\documentstyle[a4p,epsfig,12pt]{article}
\renewcommand{\Huge}{\huge}
\begin{document}
\def\ssee{\mbox{$s_{\rm e^+ e^-}$}}
\def\ssgg{s_{\gamma\gamma}}
\def\sqee{\mbox{$\sqrt{s}_{\rm ee}$}}
\def\sqeeb{\sqrt{s}_{\protect\bf\rm e^+ e^-}}
\def\sqgg{\sqrt{s}_{\gamma\gamma}}
\def\gg{\gamma\gamma}
\def\Lgg{\mbox{$L_{\rm \gamma\gamma}$}}
\def\Wgg{\mbox{$W$}}
\def\Wrec{\mbox{$W_{\rm vis}$}}
\def\Wvis{\mbox{$W_{\rm vis}$}}
\def\Wecal{\mbox{$W_{\rm ECAL}$}}
\def\qmax{\mbox{$Q^2_{\rm max}$}}
\def\qmin{\mbox{$Q^2_{\rm min}$}}
\def\Qsq{\mbox{$Q^2$}}
\def\ZO{\mbox{${\rm Z}^0$}}
\def\ee{\mbox{${\rm e}^+{\rm e}^-$}}
\def\fg{\mbox{$f_{\gamma/e}$}}
\def\sgg{\sigma_{\gamma\gamma}}
\def\see{\sigma_{\mbox{\footnotesize e}^+\mbox{\footnotesize e}^-}}
\def\zav{\langle z_0 \rangle}
\def\rav{\langle r_0 \rangle}
\def\Qav{Q}
\def\sigmagg{\sigma_{\gg}}
\def\EECAL{E_{\rm ECAL}}
\def\EFD{E_{\rm FD}}
\def\ESIW{E_{\rm SiW}}
\def\PT{|\vec{P}_T |}
\def\PL{|\vec{P}_L |_{\rm ECAL}}
\def\nch{n_{\rm ch}}
\def\dpt{1/N_{\rm ev}\cdot dE_T/d\eta}
\def\Dpt{\frac{1}{N_{\rm ev}}\frac{dE_T}{d\eta}}
\newcommand{\sleq} {\raisebox{-.6ex}{${\textstyle\stackrel{<}{\sim}}$}}
\newcommand{\sgeq} {\raisebox{-.6ex}{${\textstyle\stackrel{>}{\sim}}$}}
\def\ETJET{E^{\rm jet}_T}
\def\EJET{E^{\rm jet}}
\def\PZJET{p^{\rm jet}_z}
\def\ETi{E_{T_i}}
\def\ETMIN{E^{\rm min}_T}
\def\thetamax{\theta_{\rm max}}
\def\qqbar{\mbox{q}\overline{\mbox{q}}}
\def\qbar{\overline{q}}
\def\xg{x_{\gamma}}
\def\xgp{x_{\gamma}^+}
\def\xgm{x_{\gamma}^-}
\def\xgpm{x_{\gamma}^{\pm}}
\def\etajet{\eta^{\rm jet}}
\def\phijet{\phi^{\rm jet}}
\def\Zzero{\ifmmode {{\mathrm Z}^0} \else {${\mathrm Z}^0$} \fi}
\def\ppbar{\overline{\mbox p}\mbox{p}}
\def\pupbar{^{\mbox{\tiny(}}\overline{\mbox p}^{\mbox{\tiny)}}\mbox{p}}
\def\pz{\phantom{0}}
\def\pzz{\phantom{00}}
\def\pzzz{\phantom{.00}}
\def\shat{M_{\rm jj}}
\def\cost{\cos\theta^*}
\def\csggh{\mbox{$\sigma _{\rm \gamma \gamma \rightarrow hadrons}(\Wgg)$}}
\def\dcsee{\mbox{${\rm d} \sigma _{\rm e^+ e^-}(\Wgg)/{\rm d\Wgg}$}}
\def\dW{{\rm d}\sigma _{\rm ee}/{\rm d}W}
\def\dsee{{\rm d}\sigma _{\rm ee}}
\def\csgg{\mbox{$\sigma _{\rm \gamma\gamma}(\Wgg)$}}
\def\eeggeeh{\mbox{${\rm e^+ e^- \rightarrow e^+ e^- \gamma^\ast \gamma^\ast \rightarrow e^+ e^- hadrons}$}}
\def\ggh{\mbox{${\rm \gamma \gamma \rightarrow hadrons}$}}
\def\eeeegg{\mbox{${\rm e^+ e^- \rightarrow e^+ e^- \gamma \gamma}$}}
\def\eeeeh{\mbox{${\rm e^+ e^- \rightarrow e^+ e^- hadrons}$}}
\def\pt{p_{\rm T}}
\def\dmax{\Delta\eta_{\rm max}}
\hyphenation{PHOJET}
\hyphenation{PYTHIA}
\renewcommand{\arraystretch}{1.0}
\renewcommand{\Huge}{\huge}
\begin{titlepage}
\begin{center}{\large   EUROPEAN LABORATORY FOR PARTICLE PHYSICS
}\end{center}\bigskip
\begin{flushright}
       CERN-EP/99-076   \\ 3rd June 1999
\end{flushright}
\bigskip\bigskip\bigskip\bigskip\bigskip
\begin{center}{\huge\bf  
Total Hadronic Cross-Section of 

Photon-Photon Interactions at LEP
}\end{center}\bigskip\bigskip
\begin{center}{\LARGE The OPAL Collaboration
}\end{center}\bigskip\bigskip
\bigskip\begin{center}{\large  Abstract}\end{center}
The total hadronic cross-section $\sigmagg(W)$ for 
the interaction of real photons, $\gg\rightarrow\mbox{~hadrons}$, 
is measured for $\gg$ centre-of-mass energies $10\le W \le 110$~GeV.
The cross-section is extracted from a measurement of the 
process $\ee\rightarrow\ee\gamma^*\gamma^*\rightarrow \ee+\mbox{~hadrons}$,
using a luminosity function for the photon flux together with form factors 
for extrapolating to real photons ($Q^2=0$~GeV$^2$). 
The data were taken with the OPAL detector at LEP at $\ee$
centre-of-mass energies $\sqee=161, 172$ and $183$~GeV.
The cross-section $\sigmagg(W)$ is compared
with Regge factorisation and with the energy dependence 
observed in $\gamma$p and pp 
interactions.
The data are also compared to models which predict
a faster rise of $\sigmagg(W)$ compared to $\gamma$p and pp 
interactions due to additional hard
$\gg$ interactions not present in hadronic collisions.
\bigskip\bigskip\bigskip\bigskip
\bigskip\bigskip
\begin{center}{\large
(To be submitted to European Physical Journal C)
}\end{center}
\end{titlepage}
\begin{center}{\Large        The OPAL Collaboration
}\end{center}\bigskip
\begin{center}{
G.\thinspace Abbiendi$^{  2}$,
K.\thinspace Ackerstaff$^{  8}$,
G.\thinspace Alexander$^{ 23}$,
J.\thinspace Allison$^{ 16}$,
N.\thinspace Altekamp$^{  5}$,
K.J.\thinspace Anderson$^{  9}$,
S.\thinspace Anderson$^{ 12}$,
S.\thinspace Arcelli$^{ 17}$,
S.\thinspace Asai$^{ 24}$,
S.F.\thinspace Ashby$^{  1}$
% ...
}\end{center}\bigskip
\bigskip
$^{  1}$School of Physics and Astronomy, University of Birmingham,
Birmingham B15 2TT, UK
\newline
$^{  2}$Dipartimento di Fisica dell' Universit\`a di Bologna and INFN,
I-40126 Bologna, Italy
\newline
% ...
$^{  5}$Cavendish Laboratory, Cambridge CB3 0HE, UK
\newline
$^{  6}$Ottawa-Carleton Institute for Physics,
Department of Physics, Carleton University,
Ottawa, Ontario K1S 5B6, Canada
\newline
% ...
$^{  8}$CERN, European Organisation for Particle Physics,
CH-1211 Geneva 23, Switzerland
\newline
$^{  9}$Enrico Fermi Institute and Department of Physics,
University of Chicago, Chicago IL 60637, USA
\newline
% ...
$^{ 12}$Indiana University, Department of Physics,
Swain Hall West 117, Bloomington IN 47405, USA
\newline
% ...
$^{ 16}$Department of Physics, Schuster Laboratory, The University,
Manchester M13 9PL, UK
\newline
$^{ 17}$Department of Physics, University of Maryland,
College Park, MD 20742, USA
\newline
% ...
$^{ 23}$Department of Physics and Astronomy, Tel Aviv University,
Tel Aviv 69978, Israel
\newline
$^{ 24}$International Centre for Elementary Particle Physics and
Department of Physics, University of Tokyo, Tokyo 113-0033, and
Kobe University, Kobe 657-8501, Japan
\newline
% ...
$^{  e}$ and University of Mining and Metallurgy, Cracow
\newline
% ...
\newline
\section{Introduction}
At high $\gg$ centre-of-mass energies $W=\sqrt{s}_{\gg}$, 
the total  hadronic cross-section $\sigmagg$
for the production of hadrons in the interaction of 
two real photons is expected to be dominated by interactions
where the photons have fluctuated into a hadronic state. 
Measuring the $\sqrt{s}_{\gg}$ dependence of $\sigmagg$ 
should therefore improve our understanding of the hadronic nature of
the photon and the universal high-energy behaviour of total 
hadronic cross-sections.

Before data from LEP became available,
the total hadronic $\gg$ cross-section had only been measured
for $\gg$ centre-of-mass energies $W$ below 20~GeV by
PLUTO~\cite{bib-pluto}, TPC/2$\gamma$~\cite{bib-tpc}, 
PEP/2$\gamma$~\cite{bib-pep} and
the MD1 experiment~\cite{bib-md1}, in a kinematic region where the expected high-energy
rise of the total cross-section could not have been observed.
Using LEP data taken at $\ee$ centre-of-mass energies $\sqee=130-161$~GeV,
L3~\cite{bib-l3} has demonstrated that
the total hadronic $\gg$ cross-section in the range $5\le W \le 75$~GeV 
is consistent with the universal Regge behaviour of total cross-sections.

Processes with a 
pointlike coupling of the photon to quarks are absent in hadron-hadron 
collisions. This additional hard component in photon interactions
is therefore expected to lead to a different
energy dependence of the total cross-section for photon-induced
interactions in comparison to hadron-hadron scattering.
Models~\cite{bib-pythia,bib-phojet} 
based on perturbative Quantum Chromodynamics (pQCD) 
encompass this by including additional photon interactions, usually denoted 
``direct'' and ``anomalous'', in addition to the 
interactions which are described by the Vector Meson Dominance
model (VMD). In VMD models, the
photon fluctuates into a bound state vector meson. 
In Regge models, a different
energy dependence of the cross-section can be obtained either by
universality breaking effects or 
by introducing an additional hard pomeron~\cite{bib-hardDL}.
Thus far no sign of such a different energy
dependence has been experimentally established in comparison of
the total $\gamma$p cross-section measured by the HERA 
experiments~\cite{bib-crossH1,bib-crossZ} with total pp 
cross-sections.  Any such effect would be expected to be more pronounced in
$\gg$ interactions, since here one has two photons in the initial state. 

In this paper, we present a measurement of the total hadronic $\gg$ 
cross-section in the range $10<W<110$~GeV 
using data taken by the OPAL detector at LEP
at $\sqee=161$, $172$ and $183$~GeV. 
The integrated luminosities are 9.9, 10.0 and 54.4~pb$^{-1}$, respectively.
At these energies
above the Z$^0$ resonance, hadron production is dominated
by photon-photon collisions and background from other processes, 
e.g.~$\ee$ annihilation, is expected to be small.
The photon-photon events are selected by a series of cuts
intended to exclude backgrounds, especially from
the $\ee$ annihilation and $\gg\to\ell^+\ell^-$ 
channels ($\ell\in\{\mbox{e},\mu,\tau\}$).
In addition, an anti-tagging
condition is applied, requiring that no scattered electron\footnote{Positrons
are also referred to as electrons} was detected.
Most of the photons therefore carry only
a small negative four-momentum squared, $Q^2$, and can be considered to be
quasi-real ($Q^2 \approx 0$~GeV$^2$).
The differential cross-section ${\rm d}{\sigma}/{\rm d}W$
is measured for the process
$\ee\rightarrow\ee+\mbox{~hadrons}$, where $W$ is the invariant
mass of the hadronic system.
From this cross-section the total hadronic cross-section
$\sigma_{\gg}(W)$ for
the interaction of real photons, $\gg\rightarrow\mbox{~hadrons}$,
is extracted as a function of $W$, using a luminosity function
for the photon flux and form factors for the extrapolation to
$Q^2=0$~GeV$^2$.

\section{The OPAL detector}
A detailed description of the OPAL detector
can be found in Ref.~\cite{opaltechnicalpaper}, and
therefore only a brief account of the main features relevant
to the present analysis will be given here.
 
The central tracking system, covering the polar angle
range $|\cos\theta|<0.73$, is located inside 
a solenoidal magnet which
provides a uniform  magnetic field of 0.435~T along the beam
axis\footnote{In the OPAL coordinate system 
  the $z$ axis points in the direction of the e$^-$ beam. The
  polar angle $\theta$, the azimuthal angle $\phi$
  and the radius $r$ denote the usual spherical coordinates.}.
The magnet is surrounded in the barrel region ($|\cos\theta|<0.82$)
by a lead glass electromagnetic
calorimeter (ECAL) and a hadronic sampling calorimeter (HCAL).  
Outside the HCAL, the detector is surrounded by muon
chambers. There are similar layers of detectors in the 
endcaps ($0.82<|\cos\theta|<0.98$). 
The small-angle region from 47 to 140 mrad
around the beam pipe on both sides
of the interaction point is covered by the forward detectors (FD)
and the region from 25 to 59 mrad by the silicon-tungsten luminometers (SW).
From 1996 onwards, relevant to the data presented in this paper,
the lower boundary of the SW acceptance has been  increased to 33 mrad
following the installation of a low-angle shield to protect the
central detector against synchrotron radiation
due to the increased LEP $\ee$ beam energies.
 
Starting with the innermost components, the
tracking system consists of a high precision silicon
microvertex detector, a vertex
drift chamber, a large volume jet chamber with 159 layers of axial
anode wires and a set of $z$ chambers measuring the track coordinates
along the beam direction. 
The transverse momenta $\pt$ of tracks are measured with a precision 
parametrised by
$\sigma_{\pt}/\pt=\sqrt{0.02^2+(0.0015\cdot \pt)^2}$ ($\pt$ in GeV/$c$)
in the central region. In this paper, ``transverse''
is always defined with respect to the $z$ axis.
The jet chamber also provides 
measurements of track energy loss, ${ \rm d} E/ {\rm d}x$, 
which are used for particle identification~\cite{opaltechnicalpaper}.

The barrel and endcap sections of the ECAL  are
both constructed from lead-glass blocks, with a depth of
$24.6$ radiation lengths in the barrel region and more than 
$22$ radiation lengths in the endcaps. 
The FD consist of cylindrical lead-scintillator calorimeters with a depth of   
24 radiation lengths divided azimuthally into 16 segments.  
The electromagnetic energy resolution is about
$18\%/\sqrt{E}$, where $E$ is the energy in GeV.                                  
The SW detectors~\cite{bib-siw} each consist
of 19 layers of silicon interleaved with 18
layers of tungsten, corresponding to a total of 22 radiation
lengths. Each silicon layer consists of 16 wedge
shaped silicon detectors. The electromagnetic energy resolution is about
$25\%/\sqrt{E}$ ($E$ in GeV). 

\section{Kinematics}
\label{sec-kine}
\begin{figure}[htb]
\begin{center}
\mbox{\epsfig{file=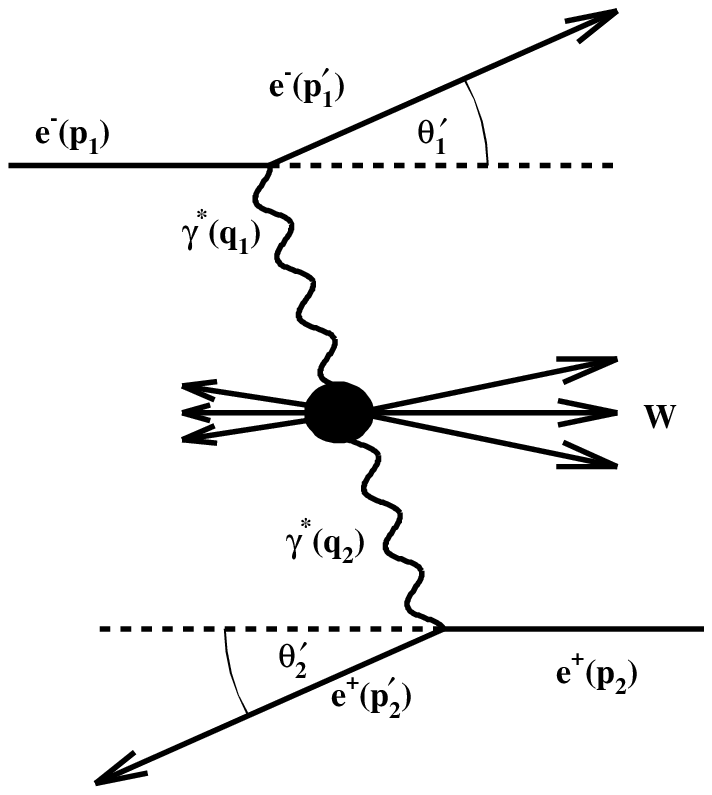,height=0.46\textwidth}}
\end{center}
\caption{Diagram of a photon-photon scattering process}
\label{fig-kine}
%
\bigskip
\epsfig{file=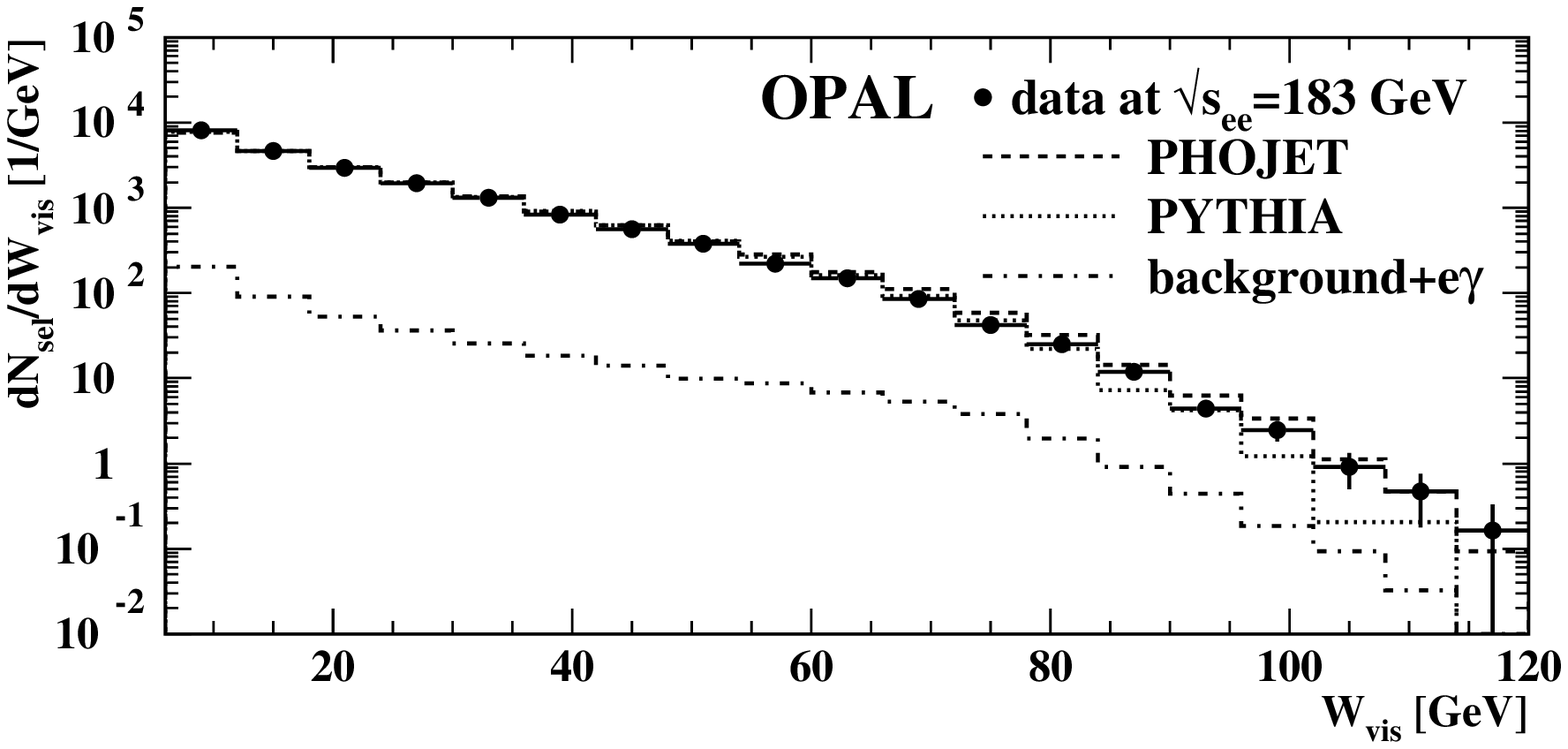,
width=1.0\textwidth,height=0.5\textwidth} 
\caption{The $\Wvis$ distribution 
for all selected events at $\sqee= 183$~GeV with $W_{\rm vis}>6$~GeV
after background subtraction.
The data are compared 
to PHOJET (dashed line) and PYTHIA (dotted line).
Only statistical errors are shown. 
The Monte Carlo background and  e$\gamma$ events which have been 
subtracted from the data
are shown as the dashed-dotted histogram.
}
\label{fig-wvis}
\end{figure}

A schematic diagram of the two-photon process is shown in Fig.~\ref{fig-kine}.
The kinematics of the process $\ee\rightarrow\ee+\mbox{~hadrons}$
at a given \sqee{} can be described by the negative square of the 
four-momentum transfers, $Q_i^2=-q_i^2$, carried by the two ($i=1,2$) 
incoming virtual photons ($\gamma^*$) and by the square of the invariant 
mass of the hadronic final 
state, $W^2=\ssgg=(q_1+q_2)^2$. The four-momenta of the electrons
before and after the interaction are denoted by $p_i$ and $p'_i$, respectively.
Each $Q_i^2$ is related to the electron
scattering angle $\theta'_i$ relative to the beam direction by
\begin{equation}
Q_i^2=-(p_i-p'_i)^2\approx 2E_i E'_i(1-\cos\theta'_i),
\label{eq-q2}
\end{equation}
where $E_i$ and $E'_i$ are the energies
of the beam electron and the scattered electron, respectively.
Events are only included in the analysis if they do not contain
scattered electrons (either single-tagged or double-tagged events).
This anti-tagging condition defines an effective upper
limit on the values of $Q_{i}^2$ for both photons. 
This condition is either met if the scattering angle $\theta'_{i}$ of
the electron is less than 33~mrad, defined by the angle between the beam axis 
and the inner edge of the acceptance of the SW detector, or if the energy of
the scattered electron is smaller than the minimum energy of 20~GeV
required for the tagged electron in SW or 40~GeV in FD.

\section{Event selection}
Two-photon events are selected with the following set of cuts:
\begin{itemize}
\item
The visible invariant mass calculated
from the position and the energy of the clusters measured
in the ECAL has to be greater than 3 GeV.
\item
The sum of all energy deposits in the ECAL and the HCAL
has to be less than 45 GeV in order to reject $\ee$ annihilation events.
\item 
At least 2 tracks must have been found in the tracking chambers. 
A track is required to have a minimum transverse momentum
of 120 MeV/$c$, at least 20 hits in the central jet chamber,
and the innermost hit of the track 
must be inside a radius of 60 cm with respect to the $z$ axis.
The point of closest approach
to the origin in the $r\phi$ plane
must be less than 20 cm in the $z$ direction and less than
1 cm in the $r\phi$ plane.
Tracks with a momentum error larger than the momentum itself
are rejected if they have less than 80 hits.  
The number of measured
hits in the jet chamber must be more than half of the number of possible hits,
where the number of possible hits
is calculated from the polar angle $\theta$ of the track, assuming 
that the track has no curvature. 
\item
The transverse momentum of the event measured in the ECAL and the FD
has to be less than 5 GeV/$c$.
\item
No track in the event has a momentum greater than 30 GeV/$c$.
\item
To remove events with scattered electrons in the FD or in the SW
calorimeters,
the total energy sum measured in the FD has to be less than
40 GeV and the total energy sum measured in the SW
less than 20 GeV.
This cut also reduces the contamination from multihadronic 
$\ee$ annihilation events with
their thrust axis close to the beam direction.
\item
The background due to beam-gas or beam-wall interactions is
reduced by the following requirements.
The radial distance of the primary vertex from the beam axis
has to be less than 3~cm.
To estimate the $z$ position of the primary vertex
for photon-photon events with typically low multiplicity, we calculate
the error-weighted average $\zav$ of the $z$ coordinates of all 
tracks at the point of closest approach to the origin in the $r\phi$ plane.
The background due to beam-gas or beam-wall interactions is
further reduced by requiring $|\zav|<10$~cm 
and that the net charge of an event, calculated by adding the
charges of all tracks, is less or equal three.
\end{itemize}

For the remaining events we determine the visible energy $E_{\rm vis}$ and
the longitudinal component, $P_{\rm L}$,
and the transverse component, $P_{\rm T}$, of the momentum vector 
of the hadronic final state. These quantities are calculated
after a matching algorithm is applied
to the data, in order to avoid double-counting of particle momenta. 
The matching algorithm uses all the information from the 
ECAL and the HCAL, the FD and the SW
calorimeters, as well as from the tracking system.
If a calorimeter energy cluster is associated to a track, 
the cluster energy is compared to the expected energy response 
$f(\vec{p})$ of the calorimeters for the track with momentum $\vec{p}$.
To calculate the energy associated to a track, the pion mass is assumed.
The cluster is rejected if the energy of the cluster is 
less than expected from the track energy. If the cluster energy $E$ exceeds
the expected energy by more than what is expected from the resolution,
the energy of the
cluster is reduced to $E - f(\vec{p})$. In this case the track momentum 
and the reduced energy of the cluster are taken separately.
The output of the matching algorithm is an array of energies and momenta 
$(E_{\rm h} , \vec{p}_{\rm h})$ which are used to calculate the visible
invariant mass \Wvis{}:
\begin{eqnarray}
W^2_{\rm vis} & = & 
\left(\sum_{h}E_{\rm h}\right)^2-\left(\sum_{{\rm h}}\vec{p}_{\rm h}
 \right)^2 \\
              & = & E_{\rm vis}^2 - P_{\rm L}^2 - P_{\rm T}^2,
\end{eqnarray}
A cut $|P_{\rm L}/E_{\rm vis}| < 0.85$ is applied, since 
beam-gas or beam-wall events
tend to accumulate at high values of $|P_{\rm L}/E_{\rm vis}|$.

In order to reject events with only leptons in the final state,
additional requirements have to be fulfilled for events
with $n_{\rm ch}=2$, where $n_{\rm ch}$ is the number of tracks.
Each of the two tracks must have at least 
20~${\rm d} E/{\rm d} x$ hits in the central jet chamber and
the ${\mathrm d}E/{\mathrm d}x$ probability must be smaller than $10\%$
for the electron and for the muon hypothesis.
The thrust of the event, calculated in the laboratory system
from the output of the matching algorithm, has to be below 0.98. 

We use data corresponding to an integrated luminosity of
74.4~pb$^{-1}$ collected with the OPAL detector in the years 1996 and 1997.
The integrated luminosity is 9.9~pb$^{-1}$ at $\sqee=161$~GeV,
10.0~pb$^{-1}$ at $\sqee=172$~GeV and 54.4~pb$^{-1}$ at $\sqee=183$~GeV.
The error on the luminosity is less than 1\%.  
After applying the above
cuts, $23250$ events  remain at $\sqee=161$~GeV, 
$25643$ at $\sqee=172$~GeV and $144147$ at $\sqee=183$~GeV. 

\section{Monte Carlo simulation}
\label{sec-MC}
The Monte Carlo generators PYTHIA 5.722~\cite{bib-pythia}
and PHOJET 1.05c~\cite{bib-phojet} are used to simulate
photon-photon interactions with $Q_1^2$ and $Q_2^2<4.5$~GeV$^2$. 
The photon-photon generator PYTHIA is based on a model
by Schuler and Sj\"ostrand~\cite{bib-GSTSZP73}
and PHOJET has been developed by Engel~\cite{bib-phojet} based on
the Dual Parton model (DPM)~\cite{bib-dpm}.
Both generators simulate the process $(\ee\rightarrow\ee+\mbox{~hadrons})$
in two stages, firstly \eeeegg{} and then \ggh{}. 
The probability of the beam electron emitting a photon is modelled by the 
Equivalent Photon Approximation (EPA)~\cite{bib-budnev}. 

Soft processes like quasi-elastic scattering
 ($\gg\to VV$, where $V$ is a vector meson), 
single-diffractive scattering ($\gg\to VX$, where $X$ is
a low mass hadronic system) or double-diffractive scattering 
($\gg\to X_1 X_2$) are modelled by both generators.
The cross-sections are obtained by fitting
a Regge parametrisation to pp, $\ppbar$ and $\gamma$p data
and by assuming Regge factorisation, i.e.~universal
couplings of the pomeron to the
hadronic fluctuations of the photon. 
In both generators
the quasi-elastic cross-section is about $5-6\%$,
the single-diffractive cross-section about $8-12\%$
and the double-diffractive cross-section about $3-4\%$ of $\sigmagg$
for $W>10$~GeV.

The transition from soft to hard interactions is defined by the
transverse momentum of the primary produced partons. 
For the hard interactions it is assumed that the cross-section
can be factorized into parton distribution functions which
give the probability to find a parton (quark, gluon) in the photon 
and matrix elements for the hard subprocess.
All possible hard interactions of quarks, gluons and photons
are simulated using leading order (LO) matrix elements. 
As default the SaS-1D parametrisation of the parton distribution 
functions~\cite{bib-sas} is used 
in PYTHIA and the LO GRV parametrisation \cite{bib-grv}
in PHOJET. 

The fragmentation and decay of the parton final state is handled 
in both generators by the routines of JETSET 7.408 \cite{bib-pythia}.
Initial- and final-state parton radiation is included 
in the leading logarithm approximation.
Both generators
include multiple interactions of the remnants of the initial photons.

The two-photon mode of PYTHIA simulates the interactions
of real photons with $Q_1^2,Q_2^2=0$~GeV$^2$. 
The virtuality of the photons, defined
by $Q^2$, enters only through the EPA in the generation of the 
photon energy spectrum, but the electrons are scattered at zero angle.
In PHOJET the $Q^2$ suppression of the total $\gg$ cross-section
is parametrised using Generalised Vector Meson Dominance (GVMD).
The $Q^2$ dependence of the quark and gluon densities of the virtual
photon and additional $Q^2$ dependent suppression factors for 
diffractive\footnote{In this paper we refer to the sum
of quasi-elastic, single- and double-diffractive events as
diffractive events} processes are also taken into account \cite{bib-phojet}.
The $Q^2$ dependent transverse momenta of the scattered electrons 
are also simulated. 

All signal and background Monte Carlo samples 
are generated with full simulation of the OPAL detector~\cite{bib-gopal}.
They are analysed using the same reconstruction algorithms as
are applied to the data.
The background from $\ee$ annihilation events 
$\ee\rightarrow(\gamma/\Zzero)^* \rightarrow \qqbar(\gamma)$ is generated with 
PYTHIA~\cite{bib-pythia}.
The leptonic two-photon background processes 
$\ee\rightarrow\ee\tau^+\tau^-$, $\ee\rightarrow\ee\mu^+\mu^-$
and $\ee\rightarrow\ee\ee$ are
simulated with VERMASEREN~\cite{bib-vermaseren}. 
The contribution from other background processes is negligible.
The Monte Carlo simulated background is less than 1.6\% of
the total number of selected events .
This does not include beam-gas and beam-wall interactions which
are estimated to contribute about $2\%$ to the total number of
selected events.

Deep-inelastic e$\gamma$ ($=\gamma^*\gamma$) events 
are generated with HERWIG 5.9~\cite{bib-her}. 
From the Monte Carlo it is estimated that after all cuts 
about 1.5\% of the remaining events are e$\gamma$ processes with
$\max\{Q_1^2,Q_2^2\}>4.5$~GeV$^2$ for $\sqee=183$~GeV.
The rate of events where both $Q_i^2$ are larger than $4.5$~GeV$^2$
is negligible. 

\section{Unfolding of the hadronic cross-section}
In a first step, the differential cross-section $\dW$ for
the process $(\ee\rightarrow\ee+\mbox{~hadrons})$ is
obtained from the $W_{\rm vis}$ distribution. 
The measured $\Wvis$ distribution is shown
in Fig.~\ref{fig-wvis} for the data taken 
at $\sqee=183$~GeV.
The distribution, which falls smoothly over 5 orders 
of magnitude,
is well described by the Monte Carlo simulations which
have been normalized to the number of data events after
subtracting the Monte Carlo expectation for background and 
e$\gamma$ events.

\begin{figure}[htb]
\epsfig{file=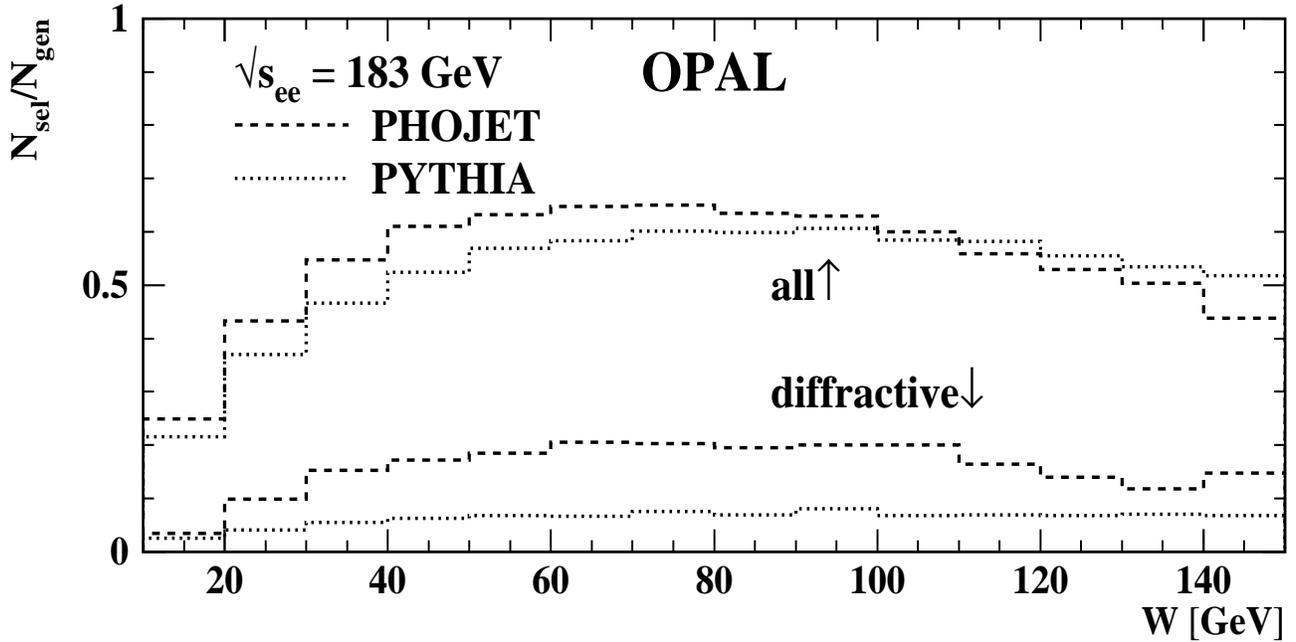,width=1.0\textwidth,height=
0.5\textwidth} 
\caption{The selection efficiency defined by
the ratio of the number of selected events, $N_{\rm sel}$, to the number
of generated events, $N_{\rm gen}$, at a given generated 
invariant mass $W$ for PHOJET (dashed line) and PYTHIA (dotted line)
at $\sqee= 183$~GeV.
The lower curves give this ratio for the sum of the diffractive 
events separately.
}
\label{fig-acc}
\end{figure}

The selection efficiency, defined as
the ratio of the number $N_{\rm sel}$ of selected Monte Carlo events 
to the number
$N_{\rm gen}$ of generated events at a given generated $W$, is
shown in Fig.~\ref{fig-acc}. It rises
from about 10\% at $W=10$~GeV to an almost
constant plateau of about $65\%$ for the PHOJET events
with $W>40$~GeV and it decreases again for very high $W$.
The selection efficiency for the PYTHIA events
is about $15\%$ lower at $W=40$~GeV and it approaches
the PHOJET selection efficiency at high $W$.
The selection efficiency for diffractive events simulated
with PHOJET is much higher than for the diffractive 
events simulated with PYTHIA.

\begin{figure}[htb]
\epsfig{file=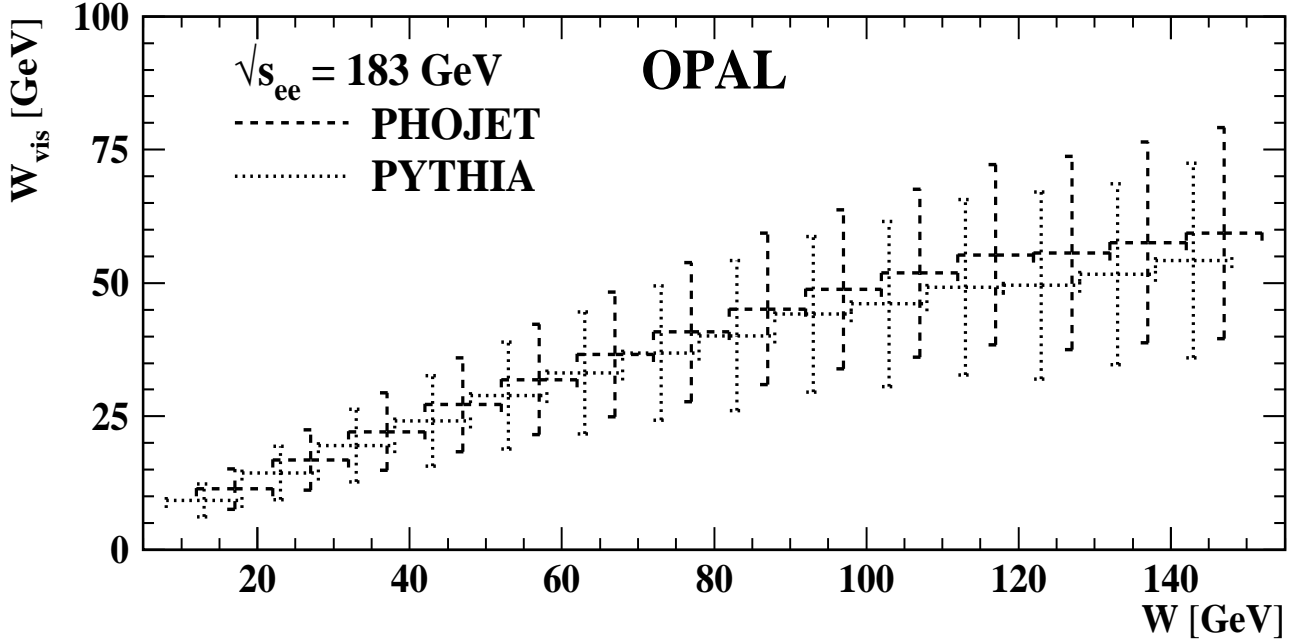,
width=1.0\textwidth,height=0.5\textwidth} 
\caption{The relation between the visible hadronic invariant mass \Wvis{}
and the generated $W$  
for all selected PHOJET (dashed) and PYTHIA (dotted)  
Monte Carlo events at $\sqee= 183$~GeV. 
The vertical bars show the standard deviation (spread) of
the $W_{\rm vis}$ distribution in each bin.
}
\label{fig-corr}
\end{figure}

The relation between \Wvis{}
and the generated $W$ for all selected PHOJET and PYTHIA Monte Carlo events
is shown in Fig.~\ref{fig-corr}. 
The finite resolution of the $W$ measurement is given
by the standard deviations of the $W_{\rm vis}$ distributions in each bin
of $W$ which are plotted as vertical bars.
The $W$ resolution is consistent for both generators, but
differences for the average \Wvis{} as a function of the generated 
$W$ are observed at high $W$.
The main energy losses are caused by
hadrons which are emitted at small polar angles $\theta$;
they are either lost in the beam pipe, or they
are only detected with low efficiency in the electromagnetic
calorimeters in the forward regions (FD and SW). 

The background determined by the Monte Carlo is first
subtracted from the data.
Then, the unfolding of the resolution effects, as well as 
the correction for the detector acceptance and the selection cuts
are done with the program GURU~\cite{bib-guru}.
This program for regularised unfolding is based on the
Singular Value Decomposition (SVD) method.
For systematic checks,
the unfolding program RUN~\cite{bib-blobel}
has also been used.
 
\begin{figure}[htb]
\epsfig{file=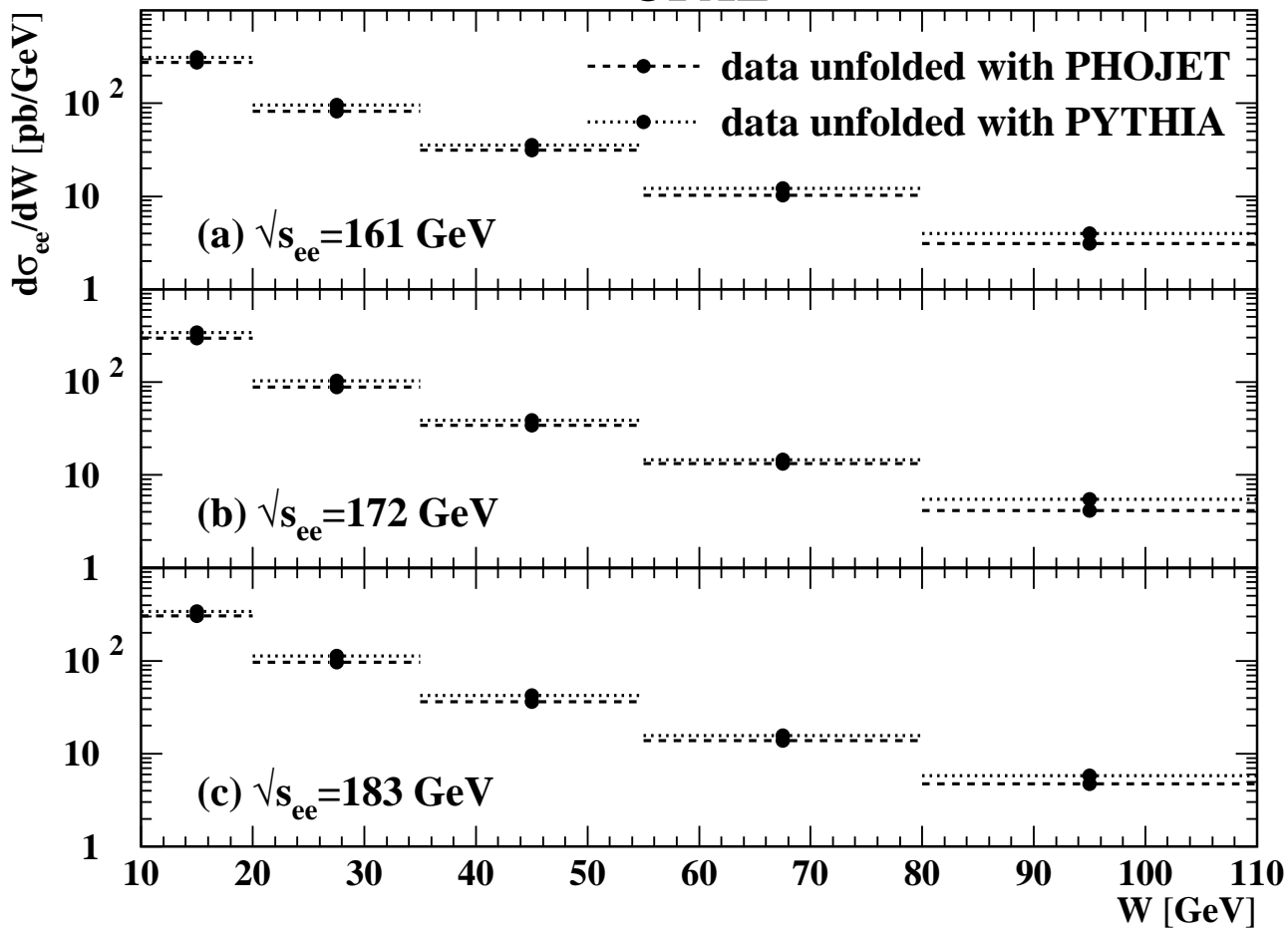,width=1.0\textwidth} 
\caption{The differential cross-section  $\dW$ at
(a) $\sqee= 161$~GeV, (b) $\sqee= 172$~GeV and (c) $\sqee= 183$~GeV
for the anti-tagged two-photon events.
Unfolding and acceptance
corrections were done with PHOJET (dashed line) and with PYTHIA (dotted line). 
The statistical errors are smaller than the symbol size.
}
\label{fig-dw}
\end{figure}

The differential cross-sections for the three beam energies, $\dW$, 
after unfolding 
are given in Fig.~\ref{fig-dw} and Table~\ref{tab-dw}. 
Bin-to-bin correlations are sizeable, since the chosen bin size 
is not much larger than the resolution. The size of the
correlations also depends on the regularisation procedure of the unfolding.
The covariance matrix obtained from the unfolding is
given in Table~\ref{tab-corr}.

The differential cross-section $\dW$ of the process 
$(\ee\rightarrow\ee+\mbox{~hadrons})$ can be translated into
the cross-section $\sigmagg$ for the process (\ggh ) using
the luminosity function $L_{\gg}$ for the photon flux~\cite{bib-budnev}.
The cross-section for real photons is derived by
using form factors $F(Q^2)$ which describe
the $Q^2$ dependence of the hadronic cross-section.
In every $W$ bin $\Delta W_i$ we determine 
\begin{equation}
\sigmagg(W'_i)=
\int_{\Delta W_i}\frac{\dsee}{{\rm d} W} {\rm d} W
\left/ \int_{\Delta W_i}\frac{{\rm d}}{{\rm d} W} \right.
\left(\int \frac{{\rm d}^4 L_{\gg}}
{{\rm d} y_1 {\rm d} Q_{1}^{2}{\rm d} y_2 {\rm d} Q_{2}^{2}} 
F(Q^2_1)F(Q^2_2)
{{\rm d} y_1 {\rm d} Q_{1}^{2}{\rm d} y_2 {\rm d} Q_{2}^{2}}\right) {\rm d} W
\label{eq-xs1}
\end{equation} 
where $y_1$ and $y_2$ denote the fraction of the beam energy
carried by the photons with $y_1y_2\approx W^2/s_{\rm ee}$ 
(neglecting $Q_1^2$ and $Q_2^2$).
The cross-section $\sigma_{\gg}$ is given at the bin centre,
since the deviation of $W'_i$ from the bin centre due to the finite
bin width is found to be small.

The luminosity function \Lgg{} and the form factors $F(Q^2)$
for the various \Wgg{} bins are obtained
from the program PHOLUM~\cite{bib-phojet} which performs 
a numerical integration for each \Wgg{} bin
over the unmeasured phase space ($Q_1^2$, $Q_2^2$ and $y$ ranges).
PHOLUM takes into account both transverse
and longitudinally polarized photons. The form factors are
used in the GVMD approximation of Ref.~\cite{bib-ginz}.
The difference between 
the extrapolation to $Q_1^2=0$ and $Q_2^2=0$ is 
about $7\%$ of $\sigma_{\gg}$ if the GVMD model is compared to 
a simple $\rho^0$ form factor~\cite{bib-phojet}.
This uncertainty is not included in the systematic error of the
measurement.

In the analysis, the e$\gamma$ events simulated by HERWIG are subtracted
from the data and
the photon flux is therefore calculated with a cut on $\max\{Q_1^2,Q_2^2\}$.
In order to check this procedure,
the analysis was also performed without subtracting the 
Monte Carlo e$\gamma$ events. In this case, the photon flux
has to be calculated without a cut on $\max\{Q_1^2,Q_2^2\}$.
The uncertainty is estimated by comparing these procedures and by
using PYTHIA instead of HERWIG for the modelling of the e$\gamma$ background.
The resulting uncertainty on $\sigma_{\gg}$ 
is about 1\% and it is therefore neglected.

Assuming the $Q^{2}_{1}$, $Q^{2}_{2}$ and $W$ dependence of
the total hadronic cross-section for virtual photons,
$\sigma_{\gamma^{*}\gamma^{*}}(Q^{2}_{1},Q^{2}_{2},W)$,
to factorize, based on a simple GVDM ansatz, the $W$ dependence is
preserved when extrapolationg to $\sigma_{\gamma\gamma}(W)$.
Although there are some events in the tails towards higher $Q_1^2$ or $Q_2^2$ 
in the $\gg$ data of this experiment,
extending to several GeV$^2$, the bulk of the data are at very low $Q_1^2$
and $Q_2^2$, so
a significant contribution from the tail is unlikely.
The medians of the
$Q_1^2$ and $Q_2^2$ distributions
are of the order $10^{-4}$~GeV$^2$ (taken from Monte Carlo). 

Radiative corrections like multiple photon emission off
the incoming electrons are not included
in the Monte Carlo generators. They are
expected to be small~\cite{bib-rad} and the effect of
the radiative corrections should be much reduced
by using the hadronic final state to calculate the
kinematics, i.e. the hadronic invariant mass $W$, of the event. 

The three data samples at $\sqee=161$, $172$ and $183$~GeV 
were independently analysed and the results
for the total hadronic two-photon cross-section $\sigmagg$
are found to be in agreement within 1-2 standard deviations
of the statistical error. Furthermore, no systematic trend
in the $W$ dependence of $\sigmagg(W)$ is observed
as a function $\sqee$. The total
cross-sections are therefore averaged using as weight the corresponding
integrated luminosities (Tab.~\ref{tab-ggxs}).

\section{Systematic errors}
\label{syserr}
Several distributions of the data are compared to PYTHIA
and PHOJET after detector simulation in order to
study whether the general description of the data by
the Monte Carlo is sufficient to use the Monte Carlo for the unfolding
of the cross-section. The Monte Carlo distributions are
all normalized to the number of data events after the Monte Carlo 
expectation for background 
and e$\gamma$ events were subtracted from the data. 
Without this normalisation, using the cross-section
predicted by the Monte Carlo generators, the number of selected events
is about 10\% smaller than in the data for PHOJET and about 10\% larger
than in the data for PYTHIA.

In both Monte Carlo models about $20\%$ of the cross-section is
due to diffractive events in which
the final state hadrons go strongly forward or backward into those parts of
the detector which have the smallest acceptance.
This fraction is almost independent of $W$ for $W>10$~GeV.
The selection efficiency for the diffractive 
events is small and, although the generated rate is almost the
same in both models, different modelling of the diffractive events leads
to very different selection efficiencies.
For a $W=70$~GeV only about 6\%
of all generated diffractive events are selected
in PYTHIA, whereas about 20\% are selected in PHOJET (Fig.~\ref{fig-acc}).
The detector correction therefore 
has to rely heavily on the Monte Carlo simulation
for this class of events. 

\begin{figure}[htb]
\epsfig{file=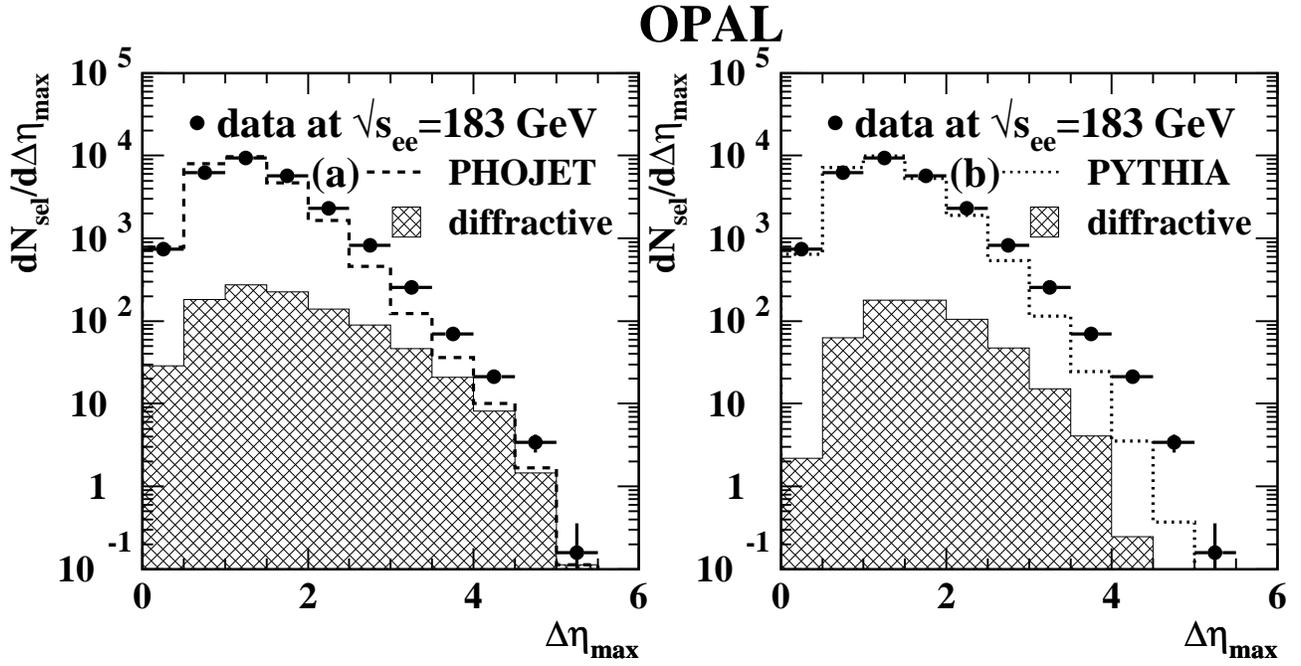,width=1.0\textwidth} 
\caption{The $\dmax$ distribution for all selected events at
$\sqee= 183$~GeV with $W_{\rm vis}>6$~GeV
after subtracting the Monte Carlo background
and e$\gamma$ events. The $\dmax$ distribution is compared
to the full a) PHOJET and b) PYTHIA simulation. The 
diffractive component is shown separately.
Only statistical errors are shown.}
\label{fig-gap}
\end{figure}

In order to study the modelling of the diffractive events
using the data, we have plotted the maximum
rapidity gap $\dmax$ between the pseudorapidities $\eta=-\ln\tan\theta/2$
of any two particles, neutral or charged, found by
the matching algorithm in Fig.~\ref{fig-gap}.
Diffractive events
are expected to have larger $\dmax$ due to the 
colour-singlet exchange~\cite{bib-bj}.
The data are compared to the PHOJET and PYTHIA simulations.
Both models underestimate the $\dmax$ distribution at large $\dmax$
with PHOJET being closer to the data than PYTHIA. 
It was checked that the transverse momenta of the scattered
electrons, which are simulated in PHOJET but not in PYTHIA, have
only a small effect on this distribution.

\begin{figure}[htb]
\epsfig{file=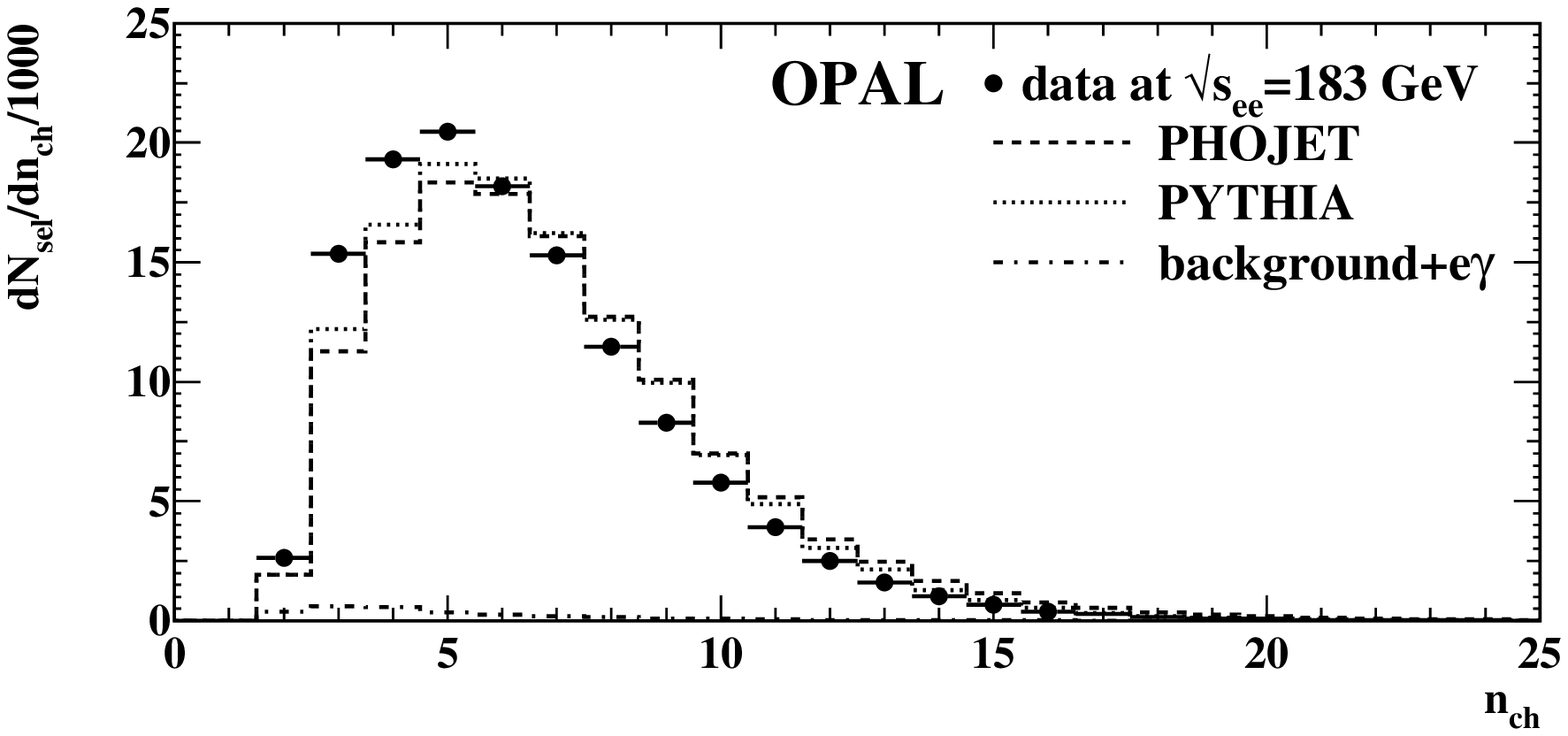,width=1.0\textwidth,height=0.5\textwidth} 
\caption{The distribution of the charged multiplicity $n_{\rm ch}$
for all selected events at $\sqee= 183$~GeV with $W_{\rm vis}>6$~GeV
compared to PHOJET (dashed line) and PYTHIA (dotted line). 
The statistical errors are smaller than the symbol size.
The subtracted Monte Carlo background and e$\gamma$ events
are shown as the dashed-dotted histogram.}
\label{fig-nch}
\end{figure}

\begin{figure}[htb]
\epsfig{file=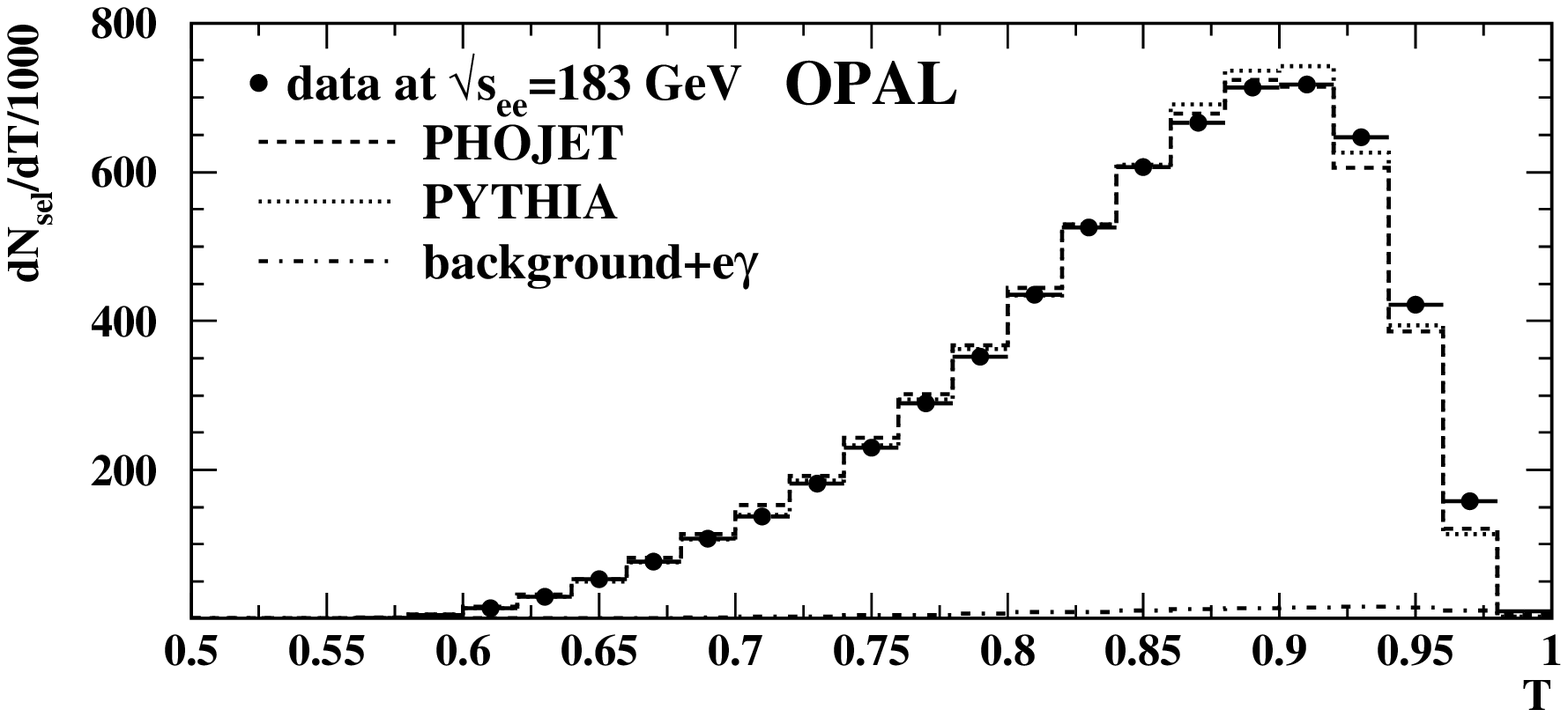,width=1.0\textwidth,height=0.5\textwidth} 
\caption{The distribution of the thrust $T$
for all selected events at $\sqee= 183$~GeV with $W_{\rm vis}>6$~GeV
compared to PHOJET (dashed line) and PYTHIA (dotted line). 
The statistical errors are smaller than the symbol size.
The subtracted Monte Carlo background and e$\gamma$ events
are shown as the dashed-dotted histogram.}
\label{fig-thrust}
\end{figure}

Significant discrepancies are also found in the distribution
of the charged multiplicity $n_{\rm ch}$ (Fig.~\ref{fig-nch}) and the
distribution of the thrust variable, $T$, (Fig.~\ref{fig-thrust}).
Both Monte Carlo models significantly underestimate
the fraction of low-multiplicity events ($n_{\rm ch}<6$) 
and overestimate the fraction of high-multiplicity events in comparison
to the data and there are also more events with large thrust ($T>0.925$), 
in the data. It should be noted that increasing the
fraction of diffractive events by factors of two or more
in the Monte Carlo
does not lead to a significant improvement in these comparisons.

\begin{figure}[htb]
\epsfig{file=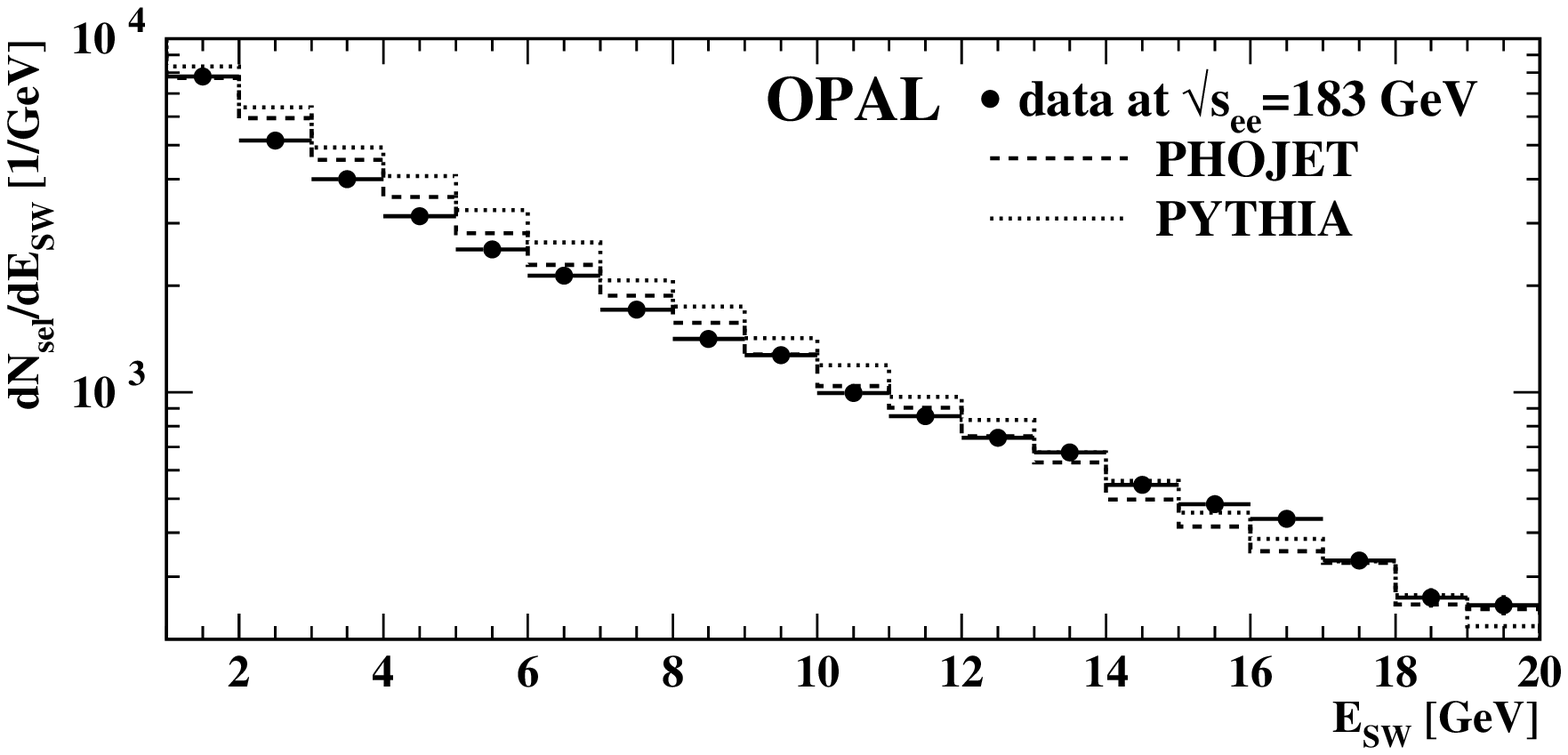,width=1.0\textwidth,height=0.5\textwidth} 
\caption{The distribution of the energy $E_{\rm SW}$ 
for all selected events at $\sqee= 183$~GeV with $W_{\rm vis}>6$~GeV
and $E_{\rm SW}>1$~GeV
compared to PHOJET (dashed line) and PYTHIA (dotted line). 
The statistical errors are smaller than the symbol size.
Due to the chosen scale
the subtracted Monte Carlo background and e$\gamma$ events
are not visible.}
\label{fig-sw}
\end{figure}

\begin{figure}[htb]
\epsfig{file=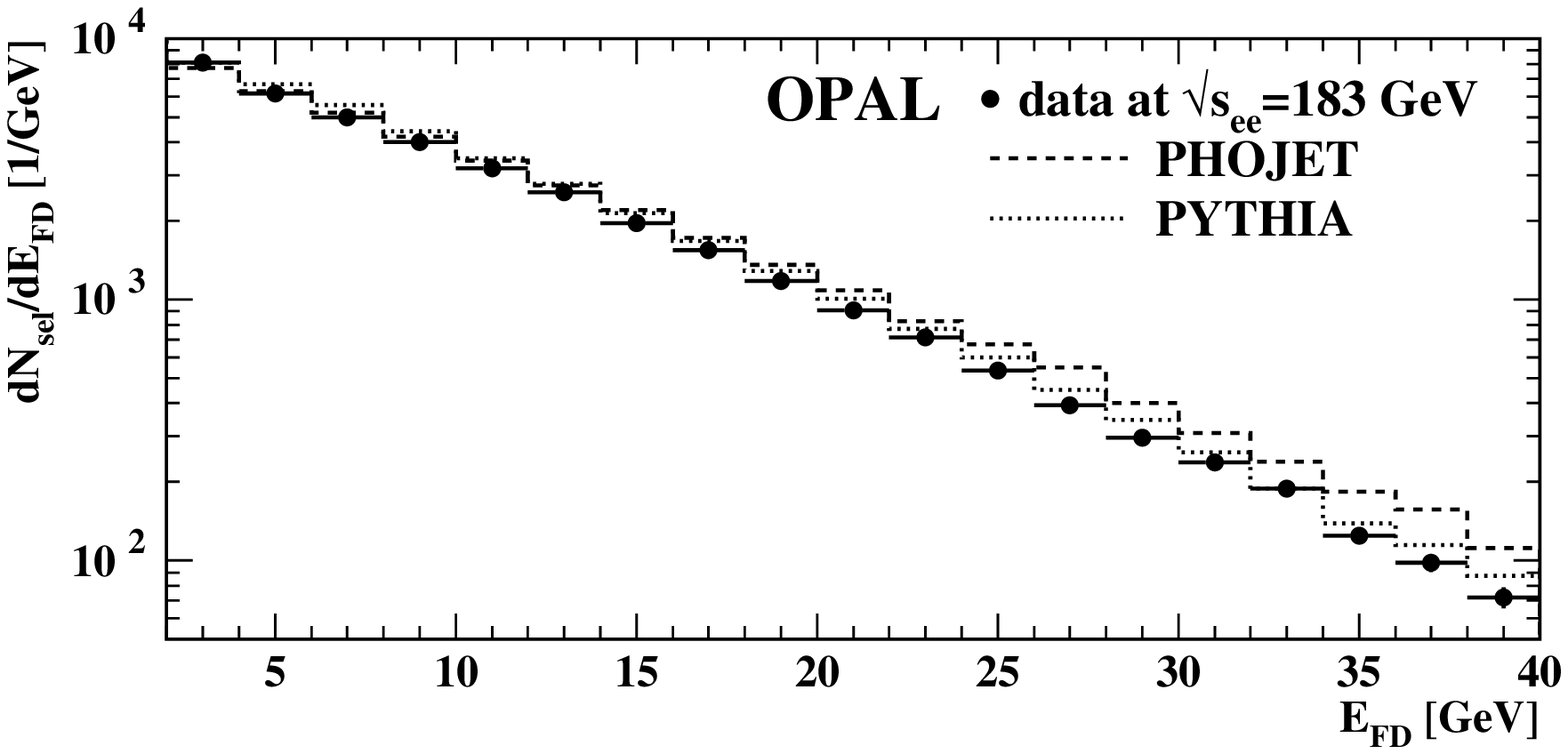,width=1.0\textwidth,height=0.5\textwidth} 
\caption{The distribution of the energy $E_{\rm FD}$ 
for all selected events at $\sqee= 183$~GeV with $W_{\rm vis}>6$~GeV
and $E_{\rm FD}>2$~GeV
compared to PHOJET (dashed line) and PYTHIA (dotted line). 
The statistical errors are smaller than the symbol size.
Due to the chosen scale
the subtracted Monte Carlo background and e$\gamma$ events
are not visible.}
\label{fig-fd}
\end{figure}

The energy $E_{\rm SW}$ measured in the silicon-tungsten luminometers (SW) 
is shown in Fig.~\ref{fig-sw} for all selected events
with $E_{\rm SW}>1$~GeV
and the energy $E_{\rm FD}$ measured in the forward detectors (FD) is 
shown in Fig.~\ref{fig-fd} for all selected events with
$E_{\rm FD}>2$~GeV. At low $E_{\rm SW}$ both Monte Carlo models
lie above the data, but 
the reasonable agreement of data and Monte Carlo
at large $E_{\rm SW}$ and $E_{\rm FD}$ shows that
the remaining background from multihadronic $\ee$ annihilation events and
deep-inelastic e$\gamma$ events is small and that
this remaining background is reasonably well described by the Monte Carlo.
This implies that there are also no events left with an off-momentum 
beam electron which was scattered upstream hitting SW or FD. 
 
\begin{figure}[htb]
\epsfig{file=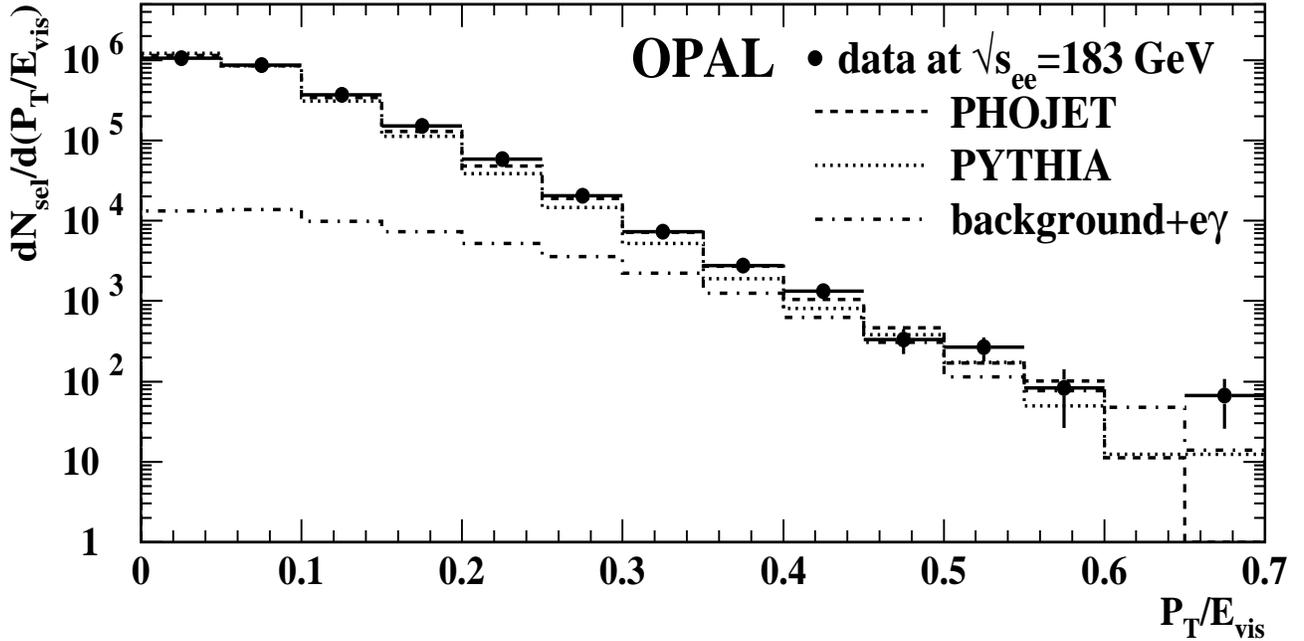,width=1.0\textwidth,height=0.5\textwidth}
\caption{The distribution of the ratio $P_{\rm T}/E_{\rm vis}$ 
for all selected events at $\sqee= 183$~GeV with $W_{\rm vis}>6$~GeV
compared to PHOJET (dashed line) and PYTHIA (dotted line). 
Only statistical errors are shown.
The subtracted Monte Carlo background and e$\gamma$ events
are shown as the dashed-dotted histogram.}
\label{fig-pt}
\end{figure}

\begin{figure}[htb]
\epsfig{file=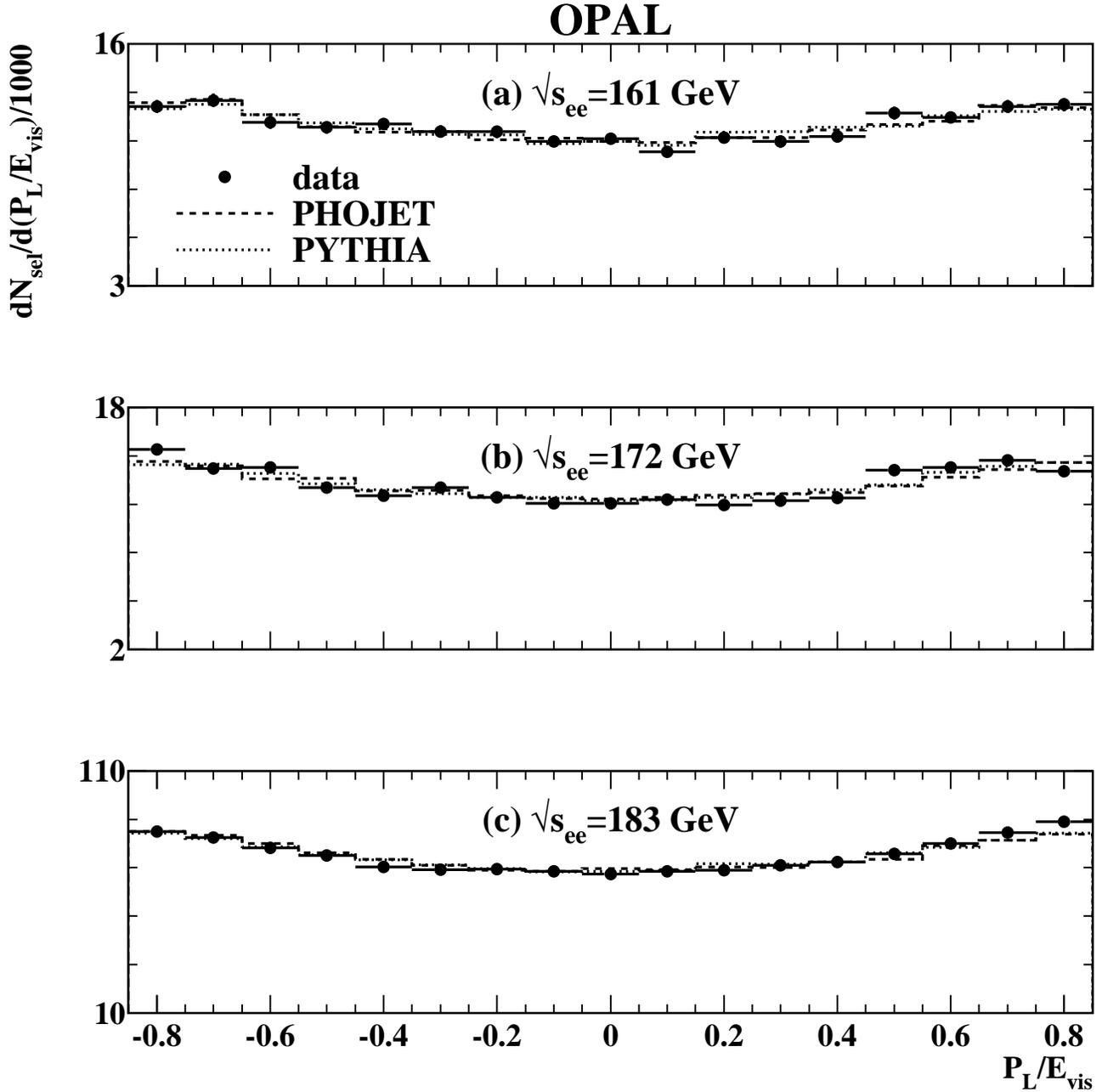,width=1.0\textwidth,height=1.0\textwidth}
\caption{The distribution of the ratio $P_{\rm L}/E_{\rm vis}$ 
for all selected events at $\sqee= 161, 172$ and $183$~GeV 
with $W_{\rm vis}>6$~GeV
compared to PHOJET (dashed line) and PYTHIA (dotted line). 
The statistical errors are smaller than the symbol size.
Due to the chosen scale
the subtracted Monte Carlo background and e$\gamma$ events
are not visible.}
\label{fig-pl}
\end{figure}

Finally, we plot the ratios $P_{\rm T}/E_{\rm vis}$
and $P_{\rm L}/E_{\rm vis}$ of the 
transverse and longitudinal components of the momentum vector of
the hadronic system to the visible total energy $E_{\rm vis}$
(Figs.~\ref{fig-pt},\ref{fig-pl}). 
Data and Monte Carlo are in good agreement.
The small number of events at large $P_{\rm T}/E_{\rm vis}$ are
expected to be mainly due to background processes.
Studies of beam-gas and beam-wall events show that most of these events
have $|P_{\rm L}/E_{\rm vis}|>0.85$. 
The background conditions were different at the three beam
energies. In the data taken at $\sqee=183$~GeV the beam
related background peaked mainly at positive $P_{\rm L}/E_{\rm vis}>0.85$.
This explains the small asymmetry in Fig.~\ref{fig-pl}c.

Based on these observations, the following systematic errors 
are taken into account in the measurement of the 
cross-sections for every beam energy, separately (Tab.~\ref{tab-dw}):
\begin{itemize}
\item
In most of the distributions, 
both Monte Carlo models describe the data equally well
and there is no reason for preferring one model over
the other for the unfolding of the data. We 
therefore average the results of the unfolding.
The difference between this cross-section 
and the results obtained by using PYTHIA or PHOJET alone
are taken as the systematic error due to the Monte Carlo model dependence
of the unfolding.
\item 
An additional error due to the uncertainties
of the modelling of the diffractive processes
in the Monte Carlo is taken into account.
Since there is large uncertainty on the diffractive $\gg$ cross-section 
derived
from the HERA measurements~\cite{bib-crossH1,bib-crossZ,bib-diff},
we have increased the percentage of diffractive events from 
$18\%$ to $27\%$ in PHOJET which leads to an increase of
$\sigmagg$ by $6\%$. 
Increasing the selection efficiency for 
diffractive events by a factor 2 leads to
a decrease of $\sigmagg$ by $6\%$. These variations
of $\pm 6\%$ are used as systematic error.
\item
The uncertainty in the ECAL energy scale was estimated to be
$\pm 3 \%$ by comparing the energy distribution reconstructed in the ECAL
for $\ee$ annihilation events at $\sqee=183$~GeV with the
Monte Carlo simulation. ECAL clusters of more than 10~GeV
were excluded from this comparison in order to have
a distribution of the energy per cluster which is similar
to $\gg$ events.
The systematic error on the total cross-section was then estimated by 
varying the reconstructed
ECAL energy in the Monte Carlo by $\pm 3 \%$.
\item
The electromagnetic calorimeters in the forward direction, SW and FD,
are used in the $W_{\rm vis}$ measurement. A possible
uncertainty in the energy scale and the detector simulation
for hadrons reconstructed in SW or FD was studied by calculating
and unfolding $W_{\rm vis}$ without SW and FD information, respectively.
The difference between $\sigma_{\rm \gg}(W)$ obtained
without SW or FD information and $\sigma_{\rm \gg}(W)$ obtained
with the full detector is taken as the systematic error.
\item
The trigger efficiency was studied using data samples which
were obtained using nearly independent sets of triggers.
The trigger efficiency is defined as the ratio of the number
of triggered and selected events to the number of selected events.
On average, the trigger efficiency for the low $W$ range, $10<W<35$~GeV,
is greater than 96\% and it approaches 100\% for larger values of $W$.
Only lower limits on the trigger efficiency can be determined with
this method and therefore no correction factor is applied. However,
the lower limit on the trigger efficiency is taken into account as an 
additional systematic error.
\item Studying vertex and net charge distributions, it
is estimated that about $2\%$ of the selected events could
be due to beam-gas or beam-wall interactions. 
Hadronic photon-photon events, however, in coincidence with an
off-momentum 
beam electron which was scattered upstream and hitting SW or FD are
rejected by the SW and FD energy cuts. 
The fraction of photon-photon events rejected due to these
coincidences is estimated to be less than $2\%$.
Taking into account both effects, a value of $2\%$
is therefore taken as additional systematic error.
\item
The program GURU~\cite{bib-guru}
has been used for unfolding the $W_{\rm vis}$ distribution.
Since the distribution of the charged multiplicity $n_{\rm ch}$
is not well described by the Monte Carlo models, we
have studied the influence of this discrepancy by
unfolding the two-dimensional $(W_{\rm vis},n_{\rm ch})$ distribution.
No significant difference from the one-dimensional unfolding
using GURU is found.
The unfolding program RUN~\cite{bib-blobel} 
can only be used for one-dimensional
unfolding.
The two unfolding methods implemented in RUN and GURU  
applied to the $W_{\rm vis}$ distribution also yield consistent results. 
Therefore no additional systematic error has been taken into account.
\item Since the background rate taken from Monte Carlo is only about $1.6\%$,
a possible systematic error is neglected.
\item
The overall normalisation error due to
the uncertainty on the luminosity measurement is less than $1\%$ and
is therefore also neglected.
\end{itemize}
The different systematic errors are summed up 
in quadrature to obtain the total systematic error.
For the total error, the statistical 
and the total systematic error are added in quadrature. 
The luminosity-weighted average values of the
total cross-section $\sigma_{\gg}(W)$
and errors for the different $W$ bins are given in Table~\ref{tab-ggxs}. 

\section{Results and model comparisons}
\label{sec-cross}
\begin{figure}[t]
   \begin{center}
      \mbox{
          \epsfxsize=1.0\textwidth
          \epsffile{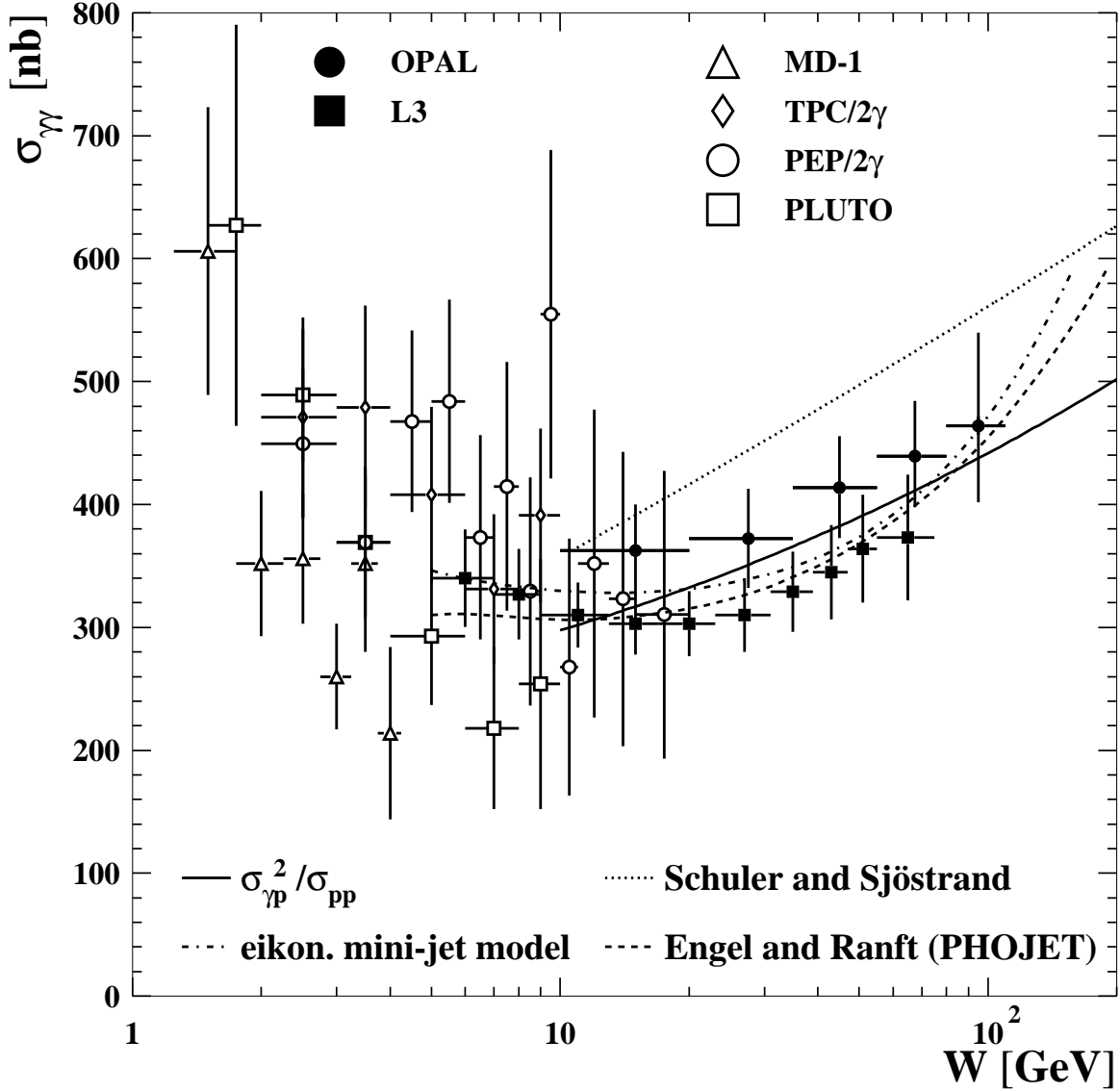}
           }
   \end{center}
\caption{
The total cross-section $\sigmagg(W)$ for the process 
$\gg\rightarrow\mbox{~hadrons}$. The OPAL measurement
is compared to measurements by
PLUTO~\cite{bib-pluto}, TPC/2$\gamma$~\cite{bib-tpc}, 
PEP/2$\gamma$~\cite{bib-pep},
MD1~\cite{bib-md1} and L3~\cite{bib-l3}. 
The statistical and the systematic errors are added in quadrature.
The normalisation uncertainty of $7\%$ due to the extrapolation
to $Q_1^2,Q_2^2=0$~GeV$^2$ is not included.
The data are compared to a simple factorisation ansatz based 
on a Donnachie-Landshoff fit to total cross-sections~\cite{bib-DL}
(solid line).
The dashed-dotted line is the eikonalised mini-jet model by Godbole and 
Panchieri~\cite{bib-minijet}, the dotted line is
the model of Schuler and
Sj\"ostrand~\cite{bib-GSTSZP73} and the dashed line is the model of Engel and 
Ranft~\cite{bib-phojet}.
}
\label{fig-ggxs}
\end{figure}

The total cross-section for the process 
$\gg\rightarrow\mbox{~hadrons}$, $\sigmagg(W)$, is shown in Fig.~\ref{fig-ggxs}
in the range $10\le W \le 110$~GeV. 
In the region $W\le 20$~GeV, the OPAL measurement is 
consistent with the results
from PLUTO~\cite{bib-pluto}, TPC/2$\gamma$~\cite{bib-tpc}
and PEP/2$\gamma$~\cite{bib-pep} within the large spread and
experimental errors of these measurements.

The OPAL measurements exhibit
the rise in the $W$ range $10<W<110$~GeV which
is characteristic for hadronic cross-sections in this energy range.
A similar rise was first observed by the L3 experiment~\cite{bib-l3},
but their values of $\sigmagg$ are about 20\% lower
than the OPAL measurement. L3 used PHOJET only for
the unfolding, whereas for the OPAL measurement presented
here the unfolding results of PHOJET and PYTHIA are
averaged. The OPAL result obtained using only PHOJET is about $5-10\%$
lower than the averaged result.

Several models have been proposed to describe the energy dependence of 
hadronic cross-sections. 
One of the interesting questions for hadronic interactions of 
real photons is 
whether they behave the same as hadrons or whether the additional 
hard contributions to the total cross-sections of photon-induced interactions
lead to a faster rise of the total $\gg$ and $\gamma$p cross-sections 
as a function of energy. Hence we performed a detailed study
of the data in the framework of present models.

We study the data within the framework of Regge theory.
The total cross-sections for hadron-hadron and photon-proton 
collisions have been found to be well described~\cite{bib-CKK97,bib-pdg98} 
by a Regge parametrisation of the form 
\begin{eqnarray}
\sigma_{\rm AB}=X_{1\rm AB} s^{\epsilon_1}+Y_{1\rm AB} s^{-\eta_1}
+Y_{2\rm AB} s^{-\eta_2}, \nonumber \\
\sigma_{\rm \bar{A}B}=X_{1\rm AB} s^{\epsilon_1}+Y_{1\rm AB} s^{-\eta_1}
-Y_{2\rm AB} s^{-\eta_2},
\label{eq-tot1}
\end{eqnarray}
where  $A$ and $B$ denote the interacting particles and the centre-of-mass 
energy squared, $s$,
is taken in units of GeV$^2$. The first term in the equation
is due to soft pomeron exchange and the other terms
are due to C-even and C-odd reggeon exchange, 
respectively~\cite{bib-pdg98}. The 
exponents
$\epsilon_1$, $\eta_1$ and $\eta_2$ are assumed to
be universal, whereas the coefficients 
$X_{1\rm AB}$ and $Y_{i\rm AB}$ are process dependent.
The values of the exponents were determined in Ref.~\cite{bib-pdg98} 
by a fit  
to the pp, $\ppbar$, $\pi^{\pm}$p, K$^{\pm}$p, $\gamma$p and $\gg$
total cross-sections:
\begin{equation}
\epsilon_1 = 0.095 \pm 0.002, \quad \quad 
\eta_1 = 0.34 \pm 0.02,
\quad {\rm and} \quad 
\eta_2 = 0.55 \pm 0.02.
\label{eq-rexp}
\end{equation}
The non-zero value of the exponent $\epsilon_1=0.095\pm0.002$ 
predicts a slow rise of the total
cross-section with energy. 
The fit in Ref.~\cite{bib-pdg98}
is dominated by the hadron-hadron data.
For $\gamma $p and $\gamma \gamma$ collisions $Y_2=0$, i.e.
Eq.~\ref{eq-tot1} reduces to the original form proposed by Donnachie and 
Landshoff~\cite{bib-DL}. In this combined fit the
available $\gg$ data, i.e. not including the OPAL data presented here,
were fitted in the range $W>4$~GeV, yielding $X_{1\gg}=(156 \pm 18$)~nb and
$Y_{\rm 1\gg}=(320\pm130)$~nb~\cite{bib-pdg98}.

Assuming factorisation of the pomeron term $X_{1\rm AB}$, 
the total $\gg$ cross-section can be related 
to the $\gamma$p and pp total cross-sections at 
centre-of-mass energies $\sqrt{s}_{\gg}=\sqrt{s}_{\rm \gamma p}=
\sqrt{s}_{\rm pp}$ larger than about 10~GeV, 
where the pomeron trajectory should dominate:
\begin{equation}
\sigma_{\gg}
\simeq \frac{\sigma_{\gamma{\rm p}}^2}{\sigma_{\rm pp }}.
\label{eq-tot2}
\end{equation}
Most models for the high-energy
behaviour of $\sigmagg$ are based on this factorisation assumption
for the soft part of the cross-section. 
In order to predict $\sigmagg$ via Eq.~\ref{eq-tot2}, the 
fit values of $X_{1\rm pp}$, $Y_{1\rm pp}$, $Y_{2\rm pp}$, 
$X_{1\rm \gamma p}$ and $Y_{1\rm \gamma p}$ 
are taken from Ref.~\cite{bib-pdg98} together with the exponents
given in Eq.~\ref{eq-rexp}. This simple factorisation
ansatz gives a reasonable description of $\sigmagg$, but a
faster increase of the cross-section than predicted by 
$\epsilon_1=0.095$ cannot be excluded, as shown in Fig.~\ref{fig-ggxs}.

The errors on the data points
are dominated by systematic errors which are highly correlated.
Therefore, the subsequent fits were made using 
the statistical error, i.e. the covariance matrix
given in Tab.~\ref{tab-corr} and the $\chi^2$ values
are only calculated with the statistical errors. 
If the fits are repeated using the correlation matrix from the unfolding 
and the total errors, the $\chi^2$ values are reduced by about 
a factor 200 but the fit results are essentially unchanged.
In order to take the correlations due to the
systematical errors fully into account, the fits were repeated
for each systematic error source, by shifting the values of the 
cross-sections at each energy by the corresponding amount of the  
systematic error. The final systematic error on the fit 
parameters is then calculated as the square root of the 
quadratic sum of the differences
between the shifted and unshifted fits to the data.

\begin{figure}[t]
   \begin{center}
      \mbox{
          \epsfxsize=1.0\textwidth
          \epsffile{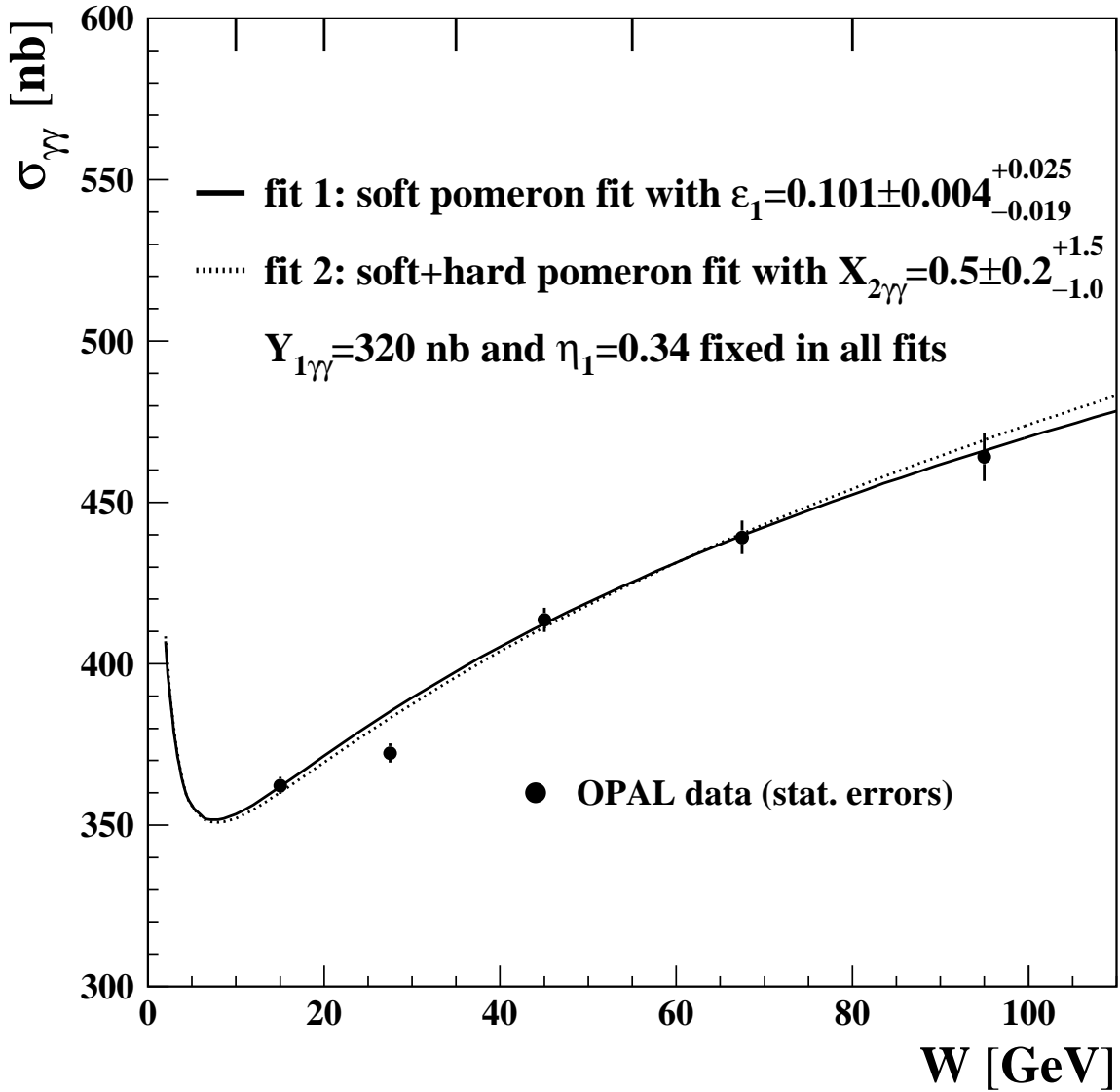}
           }
   \end{center}
\caption{
The total cross-section $\sigmagg(W)$ for the process 
$\gg\rightarrow\mbox{~hadrons}$.
Different Regge parametrisations have been fitted to the OPAL data.
Only the statistical errors are shown. The vertical lines
at the top of the figure delineate the bin boundaries.}
\label{fig-fit}
\end{figure}

In all subsequent fits to our data, we fix the reggeon term
by using the values given in Ref.~\cite{bib-pdg98},
$\eta_1=0.34$, $Y_{1\gg} = 320$~nb and $Y_{2\gg} = 0$,
since we have no data at low $W$ to constrain
the fit in this region.
We first check the universality of the exponent $\epsilon_1$
by fitting Eq.~\ref{eq-tot1} to the data, leaving
the exponent $\epsilon_1$ and the
coupling $X_{1\gg}$ free.
The results of this fit (denoted by fit 1) are given in Table~\ref{tab-fit}
and shown in Fig.~\ref{fig-fit}, together with the OPAL
data points. Only the
diagonal elements of the covariance matrix from Tab.~\ref{tab-corr}
are shown as error bars.
From fit 1 we obtain
\begin{equation}
\epsilon_1=0.101\pm0.004\mbox{(stat)}^{+0.025}_{-0.019}\mbox{(sys)}
\end{equation}
which is in agreement with the value $\epsilon_1=0.095\pm0.002$
which describes the hadron-hadron and $\gamma$p data. 

In Ref.~\cite{bib-hardDL}, a scheme is proposed for analysing 
hadron, real-photon
and virtual-photon interactions. This scheme is still based on
Regge phenomenology, but
introduces an extra term which can be identified with an additional hard 
pomeron, which has an intercept significantly larger than one.
This hard pomeron
is assumed to be responsible for the fast rise of the 
virtual photon-proton $\gamma^*$p
cross-section. In this model, the cross-section is given by
\begin{equation}
\sigma_{\rm AB}=X_{\rm 1AB} s^{\epsilon_1}
+X_{2\rm AB} s^{\epsilon_2}
+Y_{1\rm AB} s^{-\eta_1}.\\
\label{eq-2pom}
\end{equation}
In a second fit to our data (denoted by fit 2), we fixed
$\epsilon_1$ and $\eta_1$ to the values in Eq.~\ref{eq-rexp} 
and $\epsilon_2$ to the value 0.418 proposed in Ref.~\cite{bib-hardDL}
to study the significance of the term $X_2$ in Eq.~\ref{eq-2pom}. 
We obtain
\begin{equation}
X_{2\gg} = (0.5 \pm 0.2  (\mbox{stat}) ^{+1.5}_{-1.0} (\mbox{sys}))\mbox{~nb.}
\label{eq-rexp4}
\end{equation}
Hence, we find that within the precision of our data this
additional term is not required.

Total cross-sections are also described by QCD inspired models.
The total cross-sections obtained with some of 
the models are shown in Fig.~\ref{fig-ggxs}.
All these models predict
a steeper rise of $\sigmagg(W)$ than the simple factorisation ansatz
based on the pp and $\gamma$p cross-sections which is
not observed within the uncertainty of our data.
Schuler and Sj\"ostrand~\cite{bib-GSTSZP73} give a 
total cross-section for the sum of all possible event
classes in their model of $\gg$ scattering where the photon
has a direct, an anomalous and a Vector Meson Dominance (VMD) component.
The direct and anomalous components lead to additional hard interactions
which are calculated to leading order in pQCD.
Schuler and Sj\"ostrand consider the spread between this prediction and
the simple factorisation ansatz as a conservative estimate
of the theoretical band of uncertainty.

We also plot the prediction of Engel and Ranft~\cite{bib-phojet}
which is implemented in PHOJET and an eikonalised  mini-jet model by
Godbole and Panchieri~\cite{bib-minijet} which uses the 
GRV parton densities of the photon and a transverse momentum
cut-off of 2~GeV$/c$ for mini-jet production. The soft part
of the cross-section is derived from $\gamma$p data.
Another eikonalized mini-jet model, which assumes simple relations
between photon and hadron-induced partonic cross-sections, also predicts
cross-sections which are lower by about 20\%~ than
the data~\cite{bib-block}.

\section{Conclusion}
We presented a measurement of the total hadronic cross-section 
$\sigmagg(W)$ for 
the interaction of two real photons, \mbox{$\gg\rightarrow\mbox{~hadrons,}$} 
in the range $10\le W \le 110$~GeV for
data taken at $\sqee=161$, $172$ and $183$~GeV.
The cross-section is in good agreement
with a simple factorisation ansatz based on $\gamma$p and
pp data.

The energy dependence of the total cross-section has been
studied using Regge model parametrisations.
We observe the high-energy rise of the total cross-section  
which is typical for hadronic interactions.
We find that previous global fits to the $\sqrt{s}$ dependence of
$\sigma_{{\rm pp}}$, $\sigma_{{\rm \gamma p}}$ 
and $\sigma_{\gg}$ give a good
description of our results.  The fit of the 
exponent in the soft pomeron term yields
$0.101\pm0.004^{+0.025}_{-0.019}$
compared with $0.095\pm0.002$ from the global fit.  
Within the uncertainty of our measurement, no indication for a faster rise
of the $\gamma\gamma$ cross-section than in hadron-hadron or
$\gamma$p interactions is observed.

Further improvements of the description of the hadronic
final state by Monte Carlo models are necessary to reduce the systematic error
of the measurement. It will also be important to 
gain a better understanding of the diffractive processes in $\gg$ scattering.

\medskip
\bigskip\bigskip\bigskip
\appendix
\par
\section*{Acknowledgements:}
\par
We thank R.~Engel for many useful discussions and for
providing the program PHOLUM.\\
We particularly wish to thank the SL Division for the efficient operation
of the LEP accelerator at all energies
 and for their continuing close cooperation with
our experimental group.  We thank our colleagues from CEA, DAPNIA/SPP,
CE-Saclay for their efforts over the years on the time-of-flight and trigger
systems which we continue to use.  In addition to the support staff at our own
institutions we are pleased to acknowledge the  \\
Department of Energy, USA, \\
National Science Foundation, USA, \\
Particle Physics and Astronomy Research Council, UK, \\
Natural Sciences and Engineering Research Council, Canada, \\
Israel Science Foundation, administered by the Israel
Academy of Science and Humanities, \\
Minerva Gesellschaft, \\
Benoziyo Center for High Energy Physics,\\
Japanese Ministry of Education, Science and Culture (the
Monbusho) and a grant under the Monbusho International
Science Research Program,\\
Japanese Society for the Promotion of Science (JSPS),\\
German Israeli Bi-national Science Foundation (GIF), \\
Bundesministerium f\"ur Bildung, Wissenschaft,
Forschung und Technologie, Germany, \\
National Research Council of Canada, \\
Research Corporation, USA,\\
Hungarian Foundation for Scientific Research, OTKA T-029328, 
T023793 and OTKA F-023259.\\

\clearpage

\renewcommand{\arraystretch}{1.13}
\begin{table}[htbp]
  \begin{center}
\begin{tabular}{|c||c|c|c|}
\hline
 $W$-range [GeV] & $\dW$ [pb/GeV]  & $\dW$ [pb/GeV]& $\dW$ [pb/GeV]  \\
           & $\sqee=161$~GeV & $\sqee=172$~GeV & $\sqee=183$~GeV \\
\hline
10 --\pz20
& $   294.0\pm5.6^{+31.0}_{-29.9}$  
& $   319.4\pm5.5^{+33.8}_{-31.9}$
& $   320.6\pm2.4^{+32.5}_{-29.9}$   \\
20 --\pz35 
& $ \pz89.0\pm1.9^{+\pz8.9}_{-\pz8.8}$  
& $ \pz96.1\pm1.9^{+  10.3}_{-  10.3}$
& $   104.4\pm0.9^{+  11.3}_{-  11.2}$\\
35 --\pz55 
& $\pz 33.5\pm0.8^{+\pz3.2}_{-\pz3.1}$
& $\pz 36.4\pm0.8^{+\pz3.5}_{-\pz3.3}$
& $\pz 39.4\pm0.4^{+\pz4.0}_{-\pz3.9}$\\  
55 --\pz80 
& $\pz 11.2\pm0.3^{+\pz1.4}_{-\pz1.1}$
& $\pz 13.9\pm0.4^{+\pz1.4}_{-\pz1.1}$
& $\pz 14.8\pm0.2^{+\pz1.5}_{-\pz1.4}$\\  
80 --110   
& $\pzz 3.5\pm0.2^{+\pz0.7}_{-\pz0.5}$
& $\pzz 4.8\pm0.2^{+\pz1.0}_{-\pz0.7}$
& $\pzz 5.3\pm0.1^{+\pz0.8}_{-\pz0.6}$\\  
  \hline     
    \end{tabular}
    \caption{The differential cross-section $\dW$ at $\sqee=161$, $172$ 
and $183$~GeV for the anti-tagged two-photon events. 
The first error is statistical and the second systematic.}
    \label{tab-dw}
  \end{center}
\end{table}
\vfill
\vspace*{\textfloatsep}
\renewcommand{\arraystretch}{1.13}
\begin{table}[hbtp]
  \begin{center}
\begin{tabular}{|c|c|c|c|c|c|c|}    
\hline
$W$-range [GeV] 
& 10 -\pz20 & 20 -\pz35 & 35 -\pz55 & 55 -\pz80 & 80 -110 \\ 
\hline
 10 -\pz20 &\pz6.75 &\pz1.75 &  -2.86 &  -2.46 &\pz0.84 \\
 20 -\pz35 &\pz1.79 &\pz8.72 &\pz3.97 &  -3.77 &  -5.68 \\
 35 -\pz55 &  -2.86 &\pz3.97 &  13.72 &  10.07 &  -6.19 \\
 55 -\pz80 &  -2.46 &  -3.77 &  10.07 &  26.85 &  20.42 \\
 80 -  110 &\pz0.84 &  -5.68 &  -6.19 &  20.42 &  54.88 \\
  \hline
    \end{tabular}
    \caption{
       The covariance matrix of the statistical errors
       on $\sigma_{\gg}(W)$ obtained from the unfolding.
       The units are nb$^2$.}
    \label{tab-corr}
  \end{center}
\end{table}
\vfill
\vspace*{\textfloatsep}
\renewcommand{\arraystretch}{1.13}
\begin{table}[htbp]
  \begin{center}
\begin{tabular}{|c||c|c|c|c|c|}
\hline
$W$-range [GeV] &
10 --\pz20 & 20 --\pz35 & 35 --\pz55 & 55 --\pz80 & 80 --\pz110 \\
\hline\hline
$\sigmagg$ [nb]&     
$362$        & $372$    & $414$      & $439$      & $464$ \\
\hline
stat.~error &    
$\pm\pz3$     & $\pm\pz3$ & $\pm\pz4$ & $\pm\pz5$ & $\pm\pz7$ \\
\hline
MC model &    
$\pm21$      & $\pm30$ & $\pm29$ & $\pm29$ & $\pm53$ \\
diffraction &     
$\pm22$ & $\pm22$ & $\pm25$ & $\pm26$ & $\pm28$ \\
ECAL &    
$\pm16$      & $\pm\pz9$  & $\pm\pz9$ & $\pm10$ & $\pm14$ \\
no FD &    
$-\pz1$      & $-\pz9$   & $-\pz5$ & $+13$ & $+41$ \\
no SW &    
$-\pz4$      & $+\pz2$   & $+\pz7$ & $+10$ & $+11$ \\
trigger &    
$+14$        & $+\pz9$    & $+\pz8$   & $+\pz7$   & $+\pz5$   \\
beam-gas &    
$\pm\pz7$    & $\pm\pz7$   & $\pm\pz8$ & $\pm\pz9$ & $\pm\pz9$ \\ 
\hline
total syst. &
$^{+37}_{-35}$ &  
$^{+40}_{-40}$ &  
$^{+42}_{-41}$ &  
$^{+45}_{-41}$ &  
$^{+75}_{-62}$ \\ \hline 
total error &    
$^{+38}_{-35}$ &  
$^{+40}_{-40}$ &  
$^{+42}_{-41}$ &  
$^{+45}_{-41}$ &  
$^{+76}_{-62}$ \\ \hline
    \end{tabular}
    \caption{The total hadronic two-photon cross-section $\sigmagg$
    and the contributions from the various systematic errors (in nb)}
    \label{tab-ggxs}
  \end{center}
\end{table}

\renewcommand{\arraystretch}{1.13}
\begin{table}[hbtp]
  \begin{center} 
\begin{tabular}{|l||l|l|l|l|l|l|l|}
\hline 
      & $X_{\rm 1\gg}$~[nb] & $\epsilon_1$ & $X_{\rm 2\gg}$~[nb] & $\epsilon_2$ & $Y_{\rm 1\gg}$~[nb] & $\eta$ & $\chi^2/$ndf \\
\hline\hline
 fit 1 &$180\pm 5^{+30}_{-32}$ & $0.101\pm0.004^{+0.025}_{-0.019}$ & $0$ & & 320  & 0.34 & 68/3 \\
\hline
 fit 2 &$182\pm 3^{+22}_{-22}$ & $0.095$ & $0.5\pm 0.2^{+1.5}_{-1.0}$ & $0.418$   & $ 320 $  & 0.34 & 65/3 \\
\hline
\hline
 PDG~\protect\cite{bib-pdg98}  
&$156\pm 18$& $0.095\pm0.002$ & $0$ &    & $320\pm130$  & $0.34\pm0.02$ &  \\
\hline
    \end{tabular}
    \caption{Results of the various fits of Regge type 
parametrisations to the total $\gg$ cross-section. 
If no error is given, the parameter was fixed in the fit.
The values of $\chi^2$ per number of degrees of freedom (ndf)
are calculated based on
the covariance matrix of the statistical errors.
The results are compared with the OPAL data in Fig.~\protect\ref{fig-fit}.
For details see text.}
    \label{tab-fit}
  \end{center}
\end{table}

\clearpage


\end{document}